\documentclass[%
reprint,
showkeys,
amsmath,amssymb,
aps,
pra,
]{revtex4-1}
\usepackage{graphics}
\usepackage{epsfig}
\usepackage{epstopdf}
\usepackage{amsmath}
\usepackage{array}
\usepackage{subfigure}
\usepackage[colorlinks=true,linkcolor=blue,citecolor=blue,      urlcolor=blue]{hyperref}
\usepackage{tikz}
\usetikzlibrary{decorations.pathmorphing, arrows.meta, decorations.markings,decorations.text}
\begin{document}
	\title{Unified formalism for the emergence of space  in non-equilibrium description}
	\author{Hassan Basari V. T.}
	\email{basari@cusat.ac.in}
	\affiliation{%
		Department of Physics, Cochin University of Science and Technology, Kochi, Kerala 682022, India
		}%
\begin{center}
\begin{abstract}

	Previous studies indicate that the expansion law in emergence of space can be derived from the first law f thermodynamics. It has been proposed a unified formulation for the expansion law applicable to a general set of gravity theories in equilibrium description. In that formulation, which is based on the first law of thermodynamics, the non-equilibrium terms are ignored. Additionally, the structure of surface degrees of freedom in that formulation deviates from the standard notion, where $N_{sur} \ne 4S$ in general theories of gravity. This motivates us to develop a new unified formulation for the expansion law by incorporating non-equilibrium terms into the first law of thermodynamics. In this work, we formulate a unified expansion law in a non-equilibrium context, utilizing the first law of thermodynamics along with the definition of an effective Misner-Sharp energy. Compared to previous generalizations of the expansion law, our formulation reconciles the basic definition of surface degrees of freedom $N_{sur} = 4S$ for a general set of gravity theories. The unified expansion law in non-equilibrium not only highlights the direct relationship between the rate of areal volume and the difference in degrees of freedom between the surface and bulk $(N_{sur}-N_{bulk})$, but also reveals an inverse correlation with the density of surface degrees of freedom. We also demonstrate that the unified expansion law is instrumental in deriving the expansion law for a general set of gravity theories, including those with higher-order curvature corrections such as $f(R)$ theories of gravity, which require a non-equilibrium description.  
	 
	   
\end{abstract}  

\maketitle
\end{center}
\section{Introduction}
The profound interplay between gravity and thermodynamics has been well established \cite{bekenstein1,bekenstein2,bardeen,Hawking1,PhysRevD.7.2850,Davies_1975,PhysRevD.15.2738,PhysRevD.14.870}. The connection between gravity and thermodynamics began with Bekenstein's introduction of black hole entropy, relating it to the horizon area \cite{bekenstein1}. He later extended this concept by generalizing the second law of thermodynamics for regions containing black holes \cite{bekenstein2}. Concurrently, Bardeen et al. formulated the four laws of black hole mechanics, which resemble the laws of thermodynamics governing ordinary macroscopic systems \cite{bardeen}. Around the same era, Hawking demonstrated that black holes formed through matter collapse emit particles with a thermal spectrum characterized by a temperature $\mathcal{T} = \kappa/2\pi$, where $\kappa$ denotes the surface gravity of the horizon \cite{Hawking1}. Remarkably, this temperature is not exclusive to black holes but is a general trait of all spacetime horizons \cite{PhysRevD.7.2850,Davies_1975,PhysRevD.15.2738}. Additionally, it is noteworthy that an accelerating observer in flat spacetime, possessing a horizon, can attribute a temperature proportional to their acceleration $a$, given by $T = a/2\pi$ (known as the Unruh temperature) \cite{PhysRevD.14.870}. These findings collectively suggest a thermodynamic underpinning within Einstein's field equations of gravity.

Later on, the Einstein field equations were derived by demanding the validity of the Clausius relation, \(\delta Q = T dS\), at a local Rindler horizon, coupled with Bekenstein's horizon area-entropy relation \cite{1995}. Here, \(\delta Q\) represents the energy flux through the local Rindler horizon, and \(T\) denotes the Unruh temperature \cite{PhysRevD.14.870} experienced by an accelerated observer near the horizon. Subsequently, Cai and Kim \cite{Cai_Kim} derived the Friedmann equations by applying a first law of the form \(-dE = TdS\) to the apparent horizon of an \((n+1)\)-dimensional FRW universe. In this context, \(S = A/4G\) represents the horizon entropy, \(T = \dfrac{1}{2\pi r_{A}}\) signifies the temperature of the horizon, and \(-dE = A(\rho + p)H r_{A} dt\) denotes the energy flux through the apparent horizon of radius \(r_{A}\). The authors extended this result to Gauss-Bonnet gravity and the more general Lovelock gravity by incorporating the corresponding entropy.

In cosmology, unlike black hole spacetime, the presence of a perfect fluid as the source introduces well-defined pressure \(p\) and energy density \(\rho\). Padmanabhan demonstrated that Einstein's equations near any spherically symmetric horizon take the form \(dE = TdS - PdV\), where \(E\) represents the horizon's energy (proportional to its radius), \(P\) is the pressure term from the gravitational source, and \(T\) is the horizon temperature, proportional to the surface gravity \cite{Padmanabhan:2002sha, PhysRevD.74.104015}. Subsequently, Akbar and Cai derived the Friedmann equations by applying the first law of thermodynamics to the apparent horizon of a Friedmann-Robertson-Walker universe. They used a unified law, \(dE = \mathcal{T}dS + WdV\) \cite{PhysRevD.75.084003}, where \(dE\) denotes the change in the Misner-Sharp energy of matter inside the apparent horizon, \(\mathcal{T} = \kappa / 2\pi\) represents the horizon's temperature, and \(W = -\dfrac{1}{2} T^{ab} h_{ab} = \dfrac{1}{2} (\rho - p)\) stands for the work density, the orthogonal projection of the energy-momentum tensor \(T^{\mu \nu}\) onto the horizon. Interestingly, these first laws exhibit the same structure as Hayward's unified first law for trapping horizons \cite{Hayward_1998,Hayward:1998ee}. It is noteworthy that for the apparent horizon, the pressure term \(-P\) in the first law \(dE = TdS - PdV\) becomes the work density \(W\), i.e., \(P = (1/2) T_{ab} h^{ab} = -W\) \cite{PhysRevD.83.024026}.

Padmanabhan derived Newton's law of gravity by unifying the equipartition law of energy for degrees of freedom at the horizon with the thermodynamic relation for entropy, \(S = E/2T\), where \(E\) represents the effective gravitational mass and \(T\) denotes the horizon temperature. These findings underscore the intimate relationship between gravity and the thermodynamics of spacetime. While conventional thermodynamics relies on macroscopic variables like pressure and temperature, which may seem insignificant at the microscopic level, they actually arise from the collective behaviour of constituent microscopic degrees of freedom. Similarly, the connection between gravity and thermodynamics suggests that variables such as metric and curvature might diminish in significance at the microscopic scale. This understanding prompts a re-evaluation of gravity as an emergent phenomenon within a pre-existing spacetime framework. Such insights illuminate the possibility of gravity emerging as a property inherent to spacetime. Further exploration of emergent gravity is detailed in the references cited \cite{Padmanabhan:2009kr,paddy2010dec,Padmanabhan2019review}.%

Taking this emergent paradigm further, Padmanabhan proposed that space itself possesses an emergent nature in the cosmological context \cite{paddy2012jun}. He posited that the expansion of the universe (the expansion of the Hubble volume) can be explained as the emergence of space with the progression of cosmic time \(t\) \cite{paddy2012jun}. It was suggested that, for an observer for whom the cosmic background radiation appears homogeneous and isotropic, the time evolution of the Universe in Einstein's gravity can be described using the equation \(dV/dt = \ell_{p}^{2} (N_{sur} - \epsilon N_{bulk})\), known as the holographic equipartition principle. Throughout this article, we refer to it as the expansion law. According to the expansion law, the emergence of space occurs to equalize the degrees of freedom (DoF) on the horizon and the DoF in the bulk enclosed by the horizon. Building on this paradigm, Padmanabhan derived the Friedmann equation from the expansion law for a flat FRW universe in (3+1) Einstein gravity \cite{paddy2012jun}. This expansion law was extended to higher-dimensional gravity theories such as \((n+1)\)-dimensional Einstein gravity, Gauss-Bonnet gravity, and more general Lovelock gravity \cite{cai,Sheykhi2013} by suitably modifying the surface degrees of freedom on the horizon surface. Further investigations along this line employing Padmanabhan's idea of the emergent space paradigm can be found in references
 \cite{Padmanabhan2019review,paddycosmologicalconstant,krishna1,krishna2,Mahith2018,Hareesh_2019,Muhsinath,basari2,Komatsu2016,PhysRevD.88.043518,Tu_2013,PhysRevD.88.084019,PhysRevD.90.124017,SHEYKHI2018118,TU2018411,PhysRevD.99.043523,Tu_2013,FARAGALI, Yuan:2016pkz,basari1,Dheepika:2022sio,Nandu2023, Krishna2022}.
 
The novelty of the expansion law idea is greatly appreciated, and the law and the form of degrees of freedom on the surface and in the bulk are generalized to different gravity theories in various ways \cite{cai,Sheykhi2013,PhysRevD.88.043518,Tu_2013,PhysRevD.88.084019}. A unified form for the expansion law and degrees of freedom for general gravity theories in equilibrium descriptions is proposed in \cite{basari2}. The unified expansion law in equilibrium is formulated using the first law of thermodynamics in equilibrium as the basic relation. This unified law is useful in deriving the expansion law in different gravity theories that have a proper definition for the corresponding Wald entropy. Additionally, the formulation justifies the use of areal volume and the Gibbons-Hawking temperature in the expansion law.

In the emergent gravity paradigm, the surface degrees of freedom on the horizon correspond to the equipartitional degrees of freedom of heat content on the horizon, given by \(E_{sur} = 2TS\) with the horizon temperature. Consequently, the number of surface degrees of freedom is \(N_{sur} = E_{sur}/(T/2)\), yielding \(N_{sur} = 4S\) \cite{paddy2010dec}. However, the unified formulation in equilibrium description necessitates a slightly different definition in more general theories of gravity \cite{basari2}. This discrepancy is expected, as previous generalizations of the expansion law to these gravity theories have encountered similar issues \cite{cai,Sheykhi2013}. One possible reason is that the unified formulation completely ignores the non-equilibrium nature of general gravity theories. It has already been shown that gravity theories like \(f(R)\) gravity, which involve higher-order curvature terms, require a non-equilibrium description \cite{AKBAR2007243,PhysRevLett.96.121301,AKBAR20067}.

It is natural to think that the slight discrepancies in the previous unified expansion law can be rectified by formulating a new unified expansion law that incorporates the non-equilibrium behavior of general theories of gravity. In this article, we present a unified expansion law in non-equilibrium by employing the first law of thermodynamics in a non-equilibrium framework, defining an effective Misner-Sharp energy.

Upon comparing our formulation with previous generalizations of the expansion law, we observe that it harmonizes with the fundamental definition of surface degrees of freedom \(N_{sur} = 4S\) across various gravity theories. Our unified expansion law in non-equilibrium not only establishes a direct relationship between the rate of areal volume and the difference in degrees of freedom between the surface and bulk \((N_{sur} - N_{bulk})\), but also reveals an inverse correlation with the density of surface degrees of freedom. This unified expansion law proves invaluable for deriving expansion laws within a broad spectrum of gravity theories, including those with higher-order terms such as \(f(R)\) theories, which necessitate a non-equilibrium description.

The paper is structured as follows: In Section \ref{section2}, we begin by deriving the unified expansion law using the first law of thermodynamics in a non-equilibrium setting, employing energy as the effective Misner-Sharp energy. Next, in Section \ref{section3}, we derive the Friedmann equations from the unified expansion law, demonstrating the consistency of our formulation. We then proceed to obtain the expansion law in various gravity theories from the unified expansion law in Section \ref{section4}. Finally, our conclusions and discussions are presented in Section \ref{conclusion}.


%
\section{Unified expansion law from first law of thermodynamics in non-equilibrium} \label{section2}
In this section, we develop a unified expansion law that incorporates the non-equilibrium properties of more general theories of gravity. To achieve this, we begin by defining an effective Misner-Sharp energy. Subsequently, we delve into the thermodynamic structure of this effective Misner-Sharp energy and state the first law of thermodynamics in a non-equilibrium framework with the effective Misner-Sharp energy. Finally, we formulate the unified expansion law using this first law of thermodynamics in a non-equilibrium description.   
\subsection{Effective Misner-Sharp energy}
It should be noted that the definition of gravitational energy in Einstein gravity is contentious, with no universally agreed-upon definition. However, for spherically symmetric spacetime, the Misner-Sharp energy \(E\) serves as a useful definition, having properties analogous to physical energy \cite{Hayward_1996}. In general gravity theories, there are various energy definitions in the literature that lead modified field equations. The definition of energy plays a crucial role in formulation of the first law of thermodynamics for general theories of gravity.

The first law of thermodynamics in Einstein gravity for the FRW universe with an asymptotic de-Sitter epoch is well-established, following an equilibrium description. Incorporating non-equilibrium terms into the first law is more rigorous. In non-equilibrium, we must consider both entropy production and energy dissipation in the universe. This consideration is necessary to obtain the exact form of the first law of thermodynamics in non-equilibrium description \cite{PhysRevLett.96.121301,AKBAR2007243,unifiedformulationTian:2014ila}.

For any minimally coupled gravity theory with higher curvature terms, the field equation can be re-expressed in the form \cite{PhysRevD.90.104042},
\begin{equation}\label{eq:field_equation}
G_{\mu\nu} \equiv R_{\mu\nu}-\frac{1}{2}Rg_{\mu\nu} = 8\pi G_{\text{eff}}T_{\mu\nu}^{(\text{eff})},
\end{equation}
where \( G_{\mu\nu} \) is the Einstein tensor, \( R_{\mu\nu} \) is the Ricci tensor, and \( g_{\mu\nu} \) is the metric of the spacetime. Here, \( G_{\text{eff}} \) is the effective coupling strength, which depends on the specific gravity model and reduces to \( G \) in the case of Einstein gravity. The effective energy-momentum tensor, \( T_{\mu\nu}^{\text{eff}} \), can be separated in the following form, 
\begin{equation}
T_{\mu\nu}^{(\text{eff})} = T_{\mu\nu}^{(m)} + T_{\mu\nu}^{(MG)},
\end{equation}
where $T_{\mu\nu}^{(m)}$ is due to matter part and $T_{\mu\nu}^{(MG)}$ is due to the modified gravity, which arises from the dependence on the higher-order curvature terms. For an FRW universe with matter following the perfect fluid model, the effective energy-momentum tensor can be expressed in a metric-independent form as
\begin{equation}
T^{\mu(\text{eff})}_{\nu}=diag(-\rho_{\text{eff}}, P_{\text{eff}},P_{\text{eff}},P_{\text{eff}})  
\end{equation}
with $ \rho_{\text{eff}}=\rho_m+\rho_{\text{MG}}, \,  P_{\text{eff}}= P_m +p_{\text{MG}} .$ Here $\rho_m$ and $P_m$ are the energy density and pressure of matter components, while the effect of higher-order curvature correction has been incorporated into the $G_{\text{eff}}, \rho_{\text{MG}}$ and $p_{\text{MG}}.$

Following from the contracted Bianchi identities, the covariant derivative of the Einstein tensor must vanish, $ \nabla_{\mu}G^{\mu}_{~\nu} =0 $. Hence the covariant derivative of the modified field equation (\ref{eq:field_equation}) will give the conservation relation on $T_{\mu\nu}^{\text{eff}}$ as
\begin{equation} \label{eq:modified_conservation}
\nabla_{\mu}(G_{\text{eff}} T^{\mu(\text{eff})}_{~\nu}) = 0.
\end{equation}

Let us now consider an (n+1) dimensional FRW universe with metric 
\begin{equation}\label{eq:metric}
ds^{2} = h_{ab}dx^{a}dx^{b} + a^{2}r^{2}d\Omega_{n-1}^{2},
\end{equation}
where $h_{ab}= diag\left[-1, a(t)^2/1-kr^2\right]$ is the two dimensional metric  
of the  $ t-r $ surface, $a(t)$ is the scale factor of  expansion, $r$ is the co-moving radial distance and $d\Omega_{n-1}$ is the metric of (n-1)-dimensional sphere with unit radius. 	The metric (\ref{eq:metric}) should have an apparent horizon, which satisfies
$h_{ab} \partial_a\tilde{r} \partial_b\tilde{r}=0,$ (where $\tilde{r}= a(t)r$). The apparent horizon radius can be identified as
\begin{equation}
\tilde{r}_A = \frac{1}{\sqrt{H^2 + \frac{k}{a^2}}},
\end{equation}
where $H$ is the Hubble parameter. 

For homogeneous and isotropic spacetime like FRW universe, the coupling strength $G_{\text{eff}},~  \rho_{\text{eff}} $ and $ P_{\text{eff}} $ only depends on  cosmic time $t$. Hence, the only non-trivial conservation relation in Eq. (\ref{eq:modified_conservation}) is in the time component, which leads to the modified continuity relation as
\begin{equation}\label{eq: Gcontiniutyeqn_non-equilibrium}
\dot{\rho}_{\text{eff}} + nH\left(\rho_{\text{eff}} +p_{\text{eff}}\right)=-\dfrac{\dot{G}_{eff}}{G_{\text{eff}}}\rho_{\text{eff}}\equiv \dot{\rho}_{\varepsilon}~~ ,
\end{equation}
where the over-dot denotes the derivative with respect to comic time, $t.$
The term on the right-hand side of the continuity equation (\ref{eq: Gcontiniutyeqn_non-equilibrium}) can be identified as the rate of change energy dissipation density over cosmic time. This dissipation occurs due to the non-equilibrium behaviour, which balances the total energy flow of the system. The dissipation energy vanishes, when the effective density \(\rho_{\text{eff}}\) and pressure \(P_{\text{eff}}\) behave like a perfect fluid. The total dissipation of energy inside the apparent horizon can be identified as
 \begin{equation}\label{eq: dissipation.energy_non-equilibrium}
 \dot{\rho}_{\varepsilon}Vdt= -\dfrac{\dot{G}_{eff}}{G_{\text{eff}}}\rho_{\text{eff}} \Omega_{n}\tilde{r}_A^{n} dt \equiv \varepsilon dt.
 \end{equation} 
The dissipation energy is useful in formulation of first law of thermodynamics consistent with the field equation (\ref{eq:field_equation}).
 
Now we can define the total energy for the effective fluid in the n-dimensional spear of radius $\tilde{r}$ as 
\begin{equation}
E = \dfrac{n(n-1)\Omega_{n}\tilde{r}^{n-2}}{16\pi G_{\text{eff}}}\left(1- h^{ab}\partial_a\tilde{r}\partial_b\tilde{r}\right).
\end{equation}
The above form of effective energy is the n-dimensional generalization of total energy in 3-dimensional spherically symmetric space  \cite{AKBAR2007243, PhysRev.136.B571}, where the $ G $ is replaced with the effective coupling strength $G_{\text{eff}}$. Here, $\left(1- h^{ab}\partial_a\tilde{r}\partial_b\tilde{r}\right) $ can be reduced to $ \tilde{r}^{2} / \tilde{r}^{2}_{A} $ using $d\tilde{r} = adr+ H\tilde{r}dt$. Then  the total energy inside the apparent horizon can be obtained as
\begin{equation}\label{eq:effective_MisnerSharp_energy}
E = \dfrac{n(n-1)\Omega_{n}\tilde{r}_{A}^{n-2}}{16\pi G_{\text{eff}}}.
\end{equation}
The above form of energy can be identified as the effective Misner-Sharp energy. It can be note that the time-time component of the modified Field equation (\ref{eq:field_equation}) implies that the total energy due to the total effective density inside the apparent horizon,  $E=\rho_{\text{eff}}V$ is equivalent to the effective Misner-Sharp energy in Eq.(\ref{eq:effective_MisnerSharp_energy}).
\subsection{Thermodynamic structure of the effective Misner-Sharp energy}
For an observer inside the apparent horizon FRW universe feels a temperature for the apparent horizon, $\mathcal{T} =  \kappa /2\pi$ \cite{Hayward:1998ee}, where $\kappa$ is the  
surface gravity given by
\begin{equation}\label{eqn:temp1}
\kappa = \frac{1}{2 \sqrt{-h}} \partial_a\left(\sqrt{-h} h_{ab} \partial_b \tilde{r} \right) = - \frac{1}{\tilde{r}_A} \left ( 1- \frac{\dot{\tilde{r}}_A}{2H\tilde{r}_A} \right).
\end{equation}
Similarly, the horizon temperature for a approximate stationary observer is used in Clausius relations on the cosmological horizon, which is defined as $T =1/(2\pi \tilde{r}_A)$ \cite{1995}.

For modified gravity theories with field equations of the form given by Eq. (\ref{eq:field_equation}), the Wald-Kodama dynamical entropy at the apparent horizon can be derived using Wald's method \cite{PhysRevD.48.R3427, Jacobson:1993} as
\begin{equation}\label{eq:entropy}
S = \frac{A}{4G_{\text{eff}}}.	
\end{equation}	
The variation in entropy can arise not only from the increase in the area of the apparent horizon but also from the time dependence of the effective coupling strength, \(G_{\text{eff}}(t)\).

Now the modified field equation (\ref{eq:field_equation}) leads to the modified Friedmann equations
\begin{equation} \label{eq:1fridmann}
r_{A}^{-2} = \dfrac{16\pi G_{\text{eff}}}{n(n-1)}\rho_{\text{eff}}
\end{equation}
and
\begin{equation}\label{eq:2fridmann}
\dot{r}_{A} = \frac{8\pi G_{\text{eff}}}{(n-1)}Hr_{A}^{3}\left(\rho_{\text{eff}} + p_{\text{eff}}\right).
\end{equation}
Comparing Eq. (\ref{eq:effective_MisnerSharp_energy}) with the modified Friedmann equations, the effective Misner-Sharp energy follows a straightforward thermodynamic relation,
\begin{equation}\label{eq:E=TS} 
E = \rho_{\text{eff}}V = \frac{(n-1)}{2}\dfrac{1}{2\pi r_{A}} \frac{A}{4G_{\text{eff}}} = \frac{(n-1)}{2} TS.
\end{equation}
Hence, we have shown a direct holographic connection between the effective Misner-Sharp energy and the total heat content of the horizon. This is a non-trivial result in general gravity theories with higher-order curvature corrections. Such a simple relation cannot be achieved in formulations where gravitational energy is treated as the usual matter energy content \cite{basari2}.
    
\subsection{ Unified first law in non-equilibrium with effective Misner-Sharp energy }
In modified gravity theories like $f(R)$ gravity, the usual first law of thermodynamics on the horizon surface with Wald entropy fails to recover the Friedmann equations \cite{Akbar:2006er}. Meantime, Eiling et al. showed that the modified gravity theories like $f(R)$ gravity have non-equilibrium features. These gravity theories need an additional entropy production term to balance the energy supply $dQ=TdS + TdS_{P}$. The change in the field equations with effective energy-momentum tensor Eq. (\ref{eq:field_equation}) can be reinterpreted as the non-equilibrium first law of thermodynamics by taking effective Misner-Sharp energy as the energy source. One can obtain the change in effective Misner-Sharp energy as \cite{unifiedformulationTian:2014ila}  
\begin{equation}
dE= A\psi+ WdV +\varepsilon dt . 
\end{equation}
This is the unified first law in non-equilibrium description.  Here $W$ is the work density, which is defined as
\begin{equation}
W=-\frac{1}{2}T^{\alpha \beta }_{eff}h_{\alpha \beta },
\end{equation}
the $ \psi $ is the energy/heat flux density on the horizon, 
\begin{equation}
\psi_{\alpha}= T^{\beta}_{\alpha}\partial_{\alpha}r_{A}+ W\partial_{\beta}r_{A}.
\end{equation}
The term  $\varepsilon dt$ represents the total energy dissipation in the effective Misner-Sharp energy due to non-equilibrium.

During the dynamics of FRW universe, the total energy change inside the apparent horizon,  $\psi + \varepsilon dt$ will be equivalent to net change in thermal energy on the horizon. Hence, the unified first law inside the apparent horizon can be expressed as 
\begin{equation}\label{eq:unifiedfirstlaw}
dE= \mathcal{T} dS+ \mathcal{T}dS_{P}+ WdV . 
\end{equation}
For a stationary observer on the horizon surface, energy flux through the horizon will be the sum of time component of energy vector $ A\psi $ and the energy dissipation 
\begin{equation}
dE_{flux}= A\psi_{t}+\varepsilon dt = -n\Omega_{n}Hr_{A}^{n}\left(\rho_{eff}+ p_{eff}\right)+ \varepsilon dt.
\end{equation}
The respective Clausius relation in non-equilibrium can be expressed as \cite{unifiedformulationTian:2014ila}
\begin{equation}\label{eq:flux}
dQ= -dE_{flux} = TdS+TdS_{P} .
\end{equation}
 
Now, the variation in gravitational entropy can be expressed as
\begin{equation}\label{key}
dS= \frac{1}{4G_{\text{eff}}} \left[ dA - \frac{\dot{G}_{\text{eff}}}{G_{\text{eff}}}Adt\right].
\end{equation} 
One can obtain the form of entropy production $ dS_{P}$ by demanding the consistency of the thermodynamic relations with the field equations (\ref{eq:1fridmann}) and (\ref{eq:2fridmann}). Both the unified first law and the modified Clausius relation will give the same amount of entropy entropy production as
\begin{equation}\label{key}
dS_{P} =\frac{(n+1)\dot{G}_{\text{eff}}A}{8G_{\text{eff}}^2}dt = \frac{(n+1)}{2}\frac{\dot{G}_{\text{eff}}}{G_{\text{eff}}}Sdt.
\end{equation} 
This shows the consistency of the formulation. Further, we define the sum of change in gravitational entropy (Wald entropy) of the system and the entropy production as
\begin{equation}\label{eq:dSbar}
d\bar{S}\equiv dS + dS_{P} = \frac{1}{4G_{\text{eff}}} \left[ dA + \frac{(n-1)}{2}\frac{\dot{G}_{\text{eff}}}{G_{\text{eff}}}Adt\right].
\end{equation}
In non-equilibrium, it can be noted that $d\bar{S}$ is appearing in place of variation of gravitational entropy, $dS$ in all the themodynamic relations in equilibrium description. 

Now, Lets take the unified first law (\ref{eq:unifiedfirstlaw}), which  can be re-expressed as
\begin{equation}\label{key}
	\rho_{\text{eff}}~dV+ V~d\rho_{\text{eff}} =\mathcal{T}d\bar{S} +\frac{\left( \rho_{\text{eff}} - p_{\text{eff}} \right)}{2}dV.
\end{equation}
Using continuity equation the above relation can be reduced to
\begin{equation}\label{eq:basic relation}
\frac{d\bar{S}}{dt}+ \frac{\varepsilon}{T}=\frac{n\Omega_{n}Hr_{A}^{n}}{T}\left(\rho_{\text{eff}}+ p_{\text{eff}}\right)
\end{equation} 
The non-equilibrium Clausius relation also be reduced to the above simple relation (\ref{eq:basic relation}). Hence the Eq. (\ref{eq:basic relation}) represents both unified expansion law and Clausius relation on the apparent horizon in FRW universe. We use this equation to formulate the unified expansion law in this work, .

\subsection{Unified expansion law in non-equilibrium}
Now we have properly defied the unified first law in non-equilibrium using the effective Misner-Sharp energy as the energy source. In this subsection, we formulate a unified expansion law from  the first law of thermodynamics in non-equilibrium. For this, we take Eq. (\ref{eq:basic relation}) and split the term $ n(\rho_{\text{eff}}+ p_{\text{eff}}) $ suitably to get the form of effective Komar energy in the relation,
\begin{equation}
\frac{d\bar{S}}{dt}+ \frac{\varepsilon}{T}=\frac{HV}{T}\left(2\rho_{\text{eff}} + (n-2)\rho_{\text{eff}}+ np_{\text{eff}}\right).
\end{equation}
Here we used the expression for the volume, $V = \Omega_{n}r_{A}^{n}$ inside the apparent horizon. Now we define the effective Komar energy as
\begin{equation}\label{key}
E_{Komar} = \frac{2\left[\left(n-2\right)\rho_{\text{eff}} + np_{\text{eff}}\right]V}{\left(n-1\right)}.
\end{equation}
The Komar energy in the above definition is almost similar to the definition used in \cite{basari2}, the only change is that the ordinary density and pressure are replaced by effective density and pressure corresponding to effective Misner-Sharp energy (\ref{eq:effective_MisnerSharp_energy}). A notable change in the definition is that the newly defined Komar energy may also depend on the specific gravity theory. This is because the effective density may include contributions from the gravitational sector, as defined earlier. The relation can now be expressed as
\begin{equation}
\frac{d\bar{S}}{dt}+ \frac{\varepsilon}{T}=(n-1)\frac{H}{T}\left(\frac{2\rho_{\text{eff}}V}{(n-1)} + \frac{E_{Komar}}{2} \right).
\end{equation}
Using the thermodynamic structure of effective Misner-Sharp energy (\ref{eq:E=TS}) we get
\begin{equation}
\frac{d\bar{S}}{dt}+ \frac{\varepsilon}{T}=(n-1)\frac{H}{T}\left(TS + \frac{E_{Komar}}{2} \right).
\end{equation}
Multiplying both sides by $4/(n-1)$ and the above relation can be reduced to
\begin{equation}\label{eq:expansion law}
\frac{4}{n-1}\left[\frac{d\bar{S}}{dt}+ \frac{\varepsilon}{T} \right]=H\left(N_{sur} -\epsilon N_{bulk} \right),
\end{equation}
where surface  and bulk degrees of freedom are defined as
\begin{equation}\label{eq:Nsur}
N_{sur}= 4S = \frac{A}{G_{\text{eff}}}
\end{equation}
and 
\begin{equation}\label{eq:Nbulk}
N_{bulk}=-\epsilon \frac{E_{\text{Komar}}}{(1/2)T}. 
\end{equation}

Now we can define the areal density of surface degrees of freedom as $ \sigma_{sur} \equiv N_{sur}/A $. Then the definition of surface degrees of freedom (\ref{eq:Nsur}) implies that the effective coupling strength $ G_{\text{eff}}$ is inversely proportional to the density of degrees of freedom on the horizon surface,
\begin{equation}\label{key}
G_{\text{eff}}=\left(\frac{N_{sur}}{A}\right)^{-1}= \left(\sigma_{sur}\right)^{-1}.
\end{equation}   
For Einstein gravity, both the density of degrees of freedom on the horizon surface and the effective coupling strength are constants, $\sigma_{sur}=\ell_{p}^{-2}$. In general theories of gravity, the density of surface degrees of freedom may varies with the progress of cosmic time.

Now using (\ref{eq:dSbar}), (\ref{eq: dissipation.energy_non-equilibrium}) and first Friedmann equation (\ref{eq:1fridmann}), the L. H. S. of the Eq. (\ref{eq:expansion law}) can be obtained as $\left(\frac{N_{sur}}{A} \right)\tilde{r}_{A}^{-1}\frac{dV}{dt}$ and the equation can be reduced to the form
\begin{equation}\label{eq:unified expansion law}
\frac{dV}{dt}=\left(\sigma_{sur}\right)^{-1} H\tilde{r}_{A}\left(N_{sur} -\epsilon N_{bulk} \right).
\end{equation}
The above relation can be identified as the unified expansion law in non-equilibrium with the degrees of freedom on the surface and in bulk as in equations $ (\ref{eq:Nsur}) $ and $ (\ref{eq:Nbulk}) $.

 A significant outcome of the above unified expansion law in non-equilibrium (\ref{eq:unified expansion law}) is its revelation of the relationship between the density of surface degrees of freedom $ \sigma_{sur} $ and the rate of emergence $dV/dt$, in addition to the disparity in degrees of freedom $(N_{sur}-N_{bulk})$. Remarkably, the emergence rate inversely correlates with the density of surface degrees of freedom on the horizon. This suggests that space emerges more rapidly in scenarios characterized by a lower density of surface degrees of freedom across general theories of gravity. This aspect is not evident in the original proposal within Einstein gravity, where the density of surface degrees of freedom remains constant with the progression of cosmic time. 
 
Here $H\tilde{r}_{A}$ only significant in non-flat universe, which is unity for $k=0$ universe. The density of surface degrees of freedom on the horizon surface $ \sigma_{sur}$ is $(1/\ell_{p})^{n-1}$ for (n+1) Einstein gravity. Hence our unified expansion law will give the Padmanabhan's expansion law in (3+1) Einstein gravity for flat universe \cite{paddy2012jun} exactly as 
\begin{equation}\label{eq:original expansion law}
\frac{dV}{dt}=\ell_{p}^{2}\left(N_{sur} -\epsilon N_{bulk} \right).
\end{equation}  
The surface degrees of freedom will reduces to $N_{sur} = 4S= A/\ell_{P}^2$ and the effective density and pressure in the bulk degrees of freedom will reduces to the ordinary density and pressure of the cosmological components. This shows the consistency of our formulation with the Padmanabhan's basic idea of expansion law.
 
One advantage of this formulation is that we have successfully defined the surface degrees of freedom as $N_{sur} = 4S$ in more general theories of gravity without compromising the consistency with the standard Friedmann equations. Furthermore, the definition of bulk degrees of freedom incorporates additional components dependent on the specific gravity theory. Investigating the gravity dependence of bulk degrees of freedom further may provide a deeper understanding of their significance and implications.

\section{Expansion law and the Friedmann equations}\label{section3}
In this section, we show the consistency of the unified expansion law in non-equilibrium by deducing the field equations from the unified expansion law. The L. H. S. of the expansion law (\ref{eq:unified expansion law}) is
\begin{equation}\label{eq:LHS}
\frac{dV}{dt} = n\Omega_{n}\tilde{r}_{A}^{n-1}\frac{d\tilde{r}_{A}}{dt} = A\dot{\tilde{r}}_{A}.
\end{equation}
The R. H. S. of the expansion law (\ref{eq:unified expansion law}) can be expressed as
\begin{align}\label{eq:RHS}
\left(\sigma_{sur}\right)^{-1} & H\tilde{r}_{A}\left(N_{sur} -\epsilon N_{bulk} \right)=\\ 
& G_{\text{eff}} A H\tilde{r}_{A}\left(\frac{1}{G_{\text{eff}}} + \frac{E_{Komar}}{AT/2}\right),
\end{align}
where we used the definition of degrees of freedom on surface (\ref{eq:Nsur}) and bulk (\ref{eq:Nbulk}) and the area $A$ taken out the bracket. 
Now, combining equations (\ref{eq:LHS})  and (\ref{eq:RHS}), using the definition of Komar energy, we get
\begin{equation}\label{key}
\frac{\dot{\tilde{r}}_{A}}{H\tilde{r}_{A}} =    1 + \frac{4G_{\text{eff}}\left[\left(n-2\right)\rho_{\text{eff}} + np_{\text{eff}}\right]V}{\left(n-1\right)AT}.
\end{equation}
Now we put the definition of temperature $T$ and $V/A$ in the above equation will leads to 
\begin{equation}\label{key}
\frac{\dot{\tilde{r}}_{A}}{H\tilde{r}_{A}^3} -  \frac{1}{\tilde{r}_{A}^2} = \frac{8\pi G_{\text{eff}}}{n\left(n-1\right)}\left[\left(n-2\right)\rho_{\text{eff}} + np_{\text{eff}}\right].
\end{equation}
Integrating the above equation using the continuity equation (\ref{eq: Gcontiniutyeqn_non-equilibrium}), after multiplying both sides by the factor $ 2\dot{a}a $ will lead to the first Friedmann equation corresponding to the field equation (\ref{eq:field_equation}),
\begin{equation}\label{eq:Friedmanneqn1}
H^{2}+\frac{k}{a^2}  = \dfrac{16\pi G_{\text{eff}}}{n(n-1)}\rho_{\text{eff}}. 
\end{equation}
The time derivative of the Friedmann equation (\ref{eq:Friedmanneqn1}) will lead to the useful relation,
\begin{equation}\label{eq:Friedmanneqn2}
\frac{\dot{\tilde{r}}_{A}}{H\tilde{r}_{A}^{3}}= \dfrac{8\pi G_{\text{eff}}}{(n-1)}\left[\rho_{\text{eff}}+ p_{\text{eff}}\right].
\end{equation}
Here we have used the continuity equation (\ref{eq: Gcontiniutyeqn_non-equilibrium}) in arriving the above relation. Equation (\ref{eq:Friedmanneqn1}) is equivalent to the second Friedmann equation corresponding to the field equation (\ref{eq:field_equation}). 
Therefore, we have demonstrated that the Friedmann equations in any minimally coupled gravity theory can be derived from the unified expansion law. This underscores the consistency of our formulation with the Friedmann equations.

Now we have formulated the unified expansion law (\ref{eq:unified expansion law}) by incorporating the non-equilibrium characteristics of more general theories of gravity, we can derive the appropriate expansion law for any minimally coupled gravity theory from this unified expansion law (\ref{eq:unified expansion law}). We'll illustrate some examples in the upcoming section. 

\section{EXPANSION LAW IN DIFFERENT THEORIES OF GRAVITY FROM THE GENERAL LAW OF EXPANSION in non-equilibrium description}\label{section4}
Now we will obtain the law of expansion in different gravity theories from our unified expansion law in non-equilibrium (\ref{eq:unified expansion law}) and we will compare the newly obtained expansion laws with the expansion laws existing in litrature.

\subsection{Gauss-Bonnet gravity}
For Gauss-Bonnet gravity the gravitational entropy has the form \cite{Sheykhi2013,PhysRevD.65.084014, PhysRevD.69.104025}
\begin{equation}\label{eq:entropy_gaussbonnet}
S= \frac{A}{4\ell_{P}^{n-1}}\left(1 + \frac{n-1}{n-3}\frac{2\tilde{\alpha}}{\tilde{r}_{A}^{2}}\right).
\end{equation}
where $\tilde{\alpha}$ is the Gauss-Bonnet coefficient which has the dimension of $\tilde{r}_{A}^{2}$.         
Correspondingly, the degrees of freedom on the horizon surface will be
\begin{equation}\label{eq:NsurGB}
N_{sur}  = \frac{A}{\ell_{P}^{n-1}}\left(1 + \frac{n-1}{n-3}\frac{2\tilde{\alpha}}{\tilde{r}_{A}^{2}}\right).
\end{equation}
Then the density of surface degrees of freedom, $ \sigma_{sur} $ for Gauss-Bonnet gravity varies with the horizon radius. Then the effective coupling strength can be obtained as	
\begin{equation}\label{eq:GBGeff}
G_{\text{eff}}=\left(\sigma_{sur}\right)^{-1} = \ell_{P}^{n-1}\left(1 + \frac{n-1}{n-3}\frac{2\tilde{\alpha}}{\tilde{r}_{A}^{2}}\right)^{-1} 
\end{equation}  
Corresponding to which the the additional density in the effective Misner-Sharp energy due to Gauss-Bonnet gravity is
\begin{equation}\label{eq:rhomg_GB}
\rho_{_{MG}} =\frac{n(n-1)(n+1) \tilde{\alpha}}{(n-3)16\pi\ell_{P}^{n-1} \tilde{r}_{A}^{4}}.
\end{equation}
Now using the continuity equation we get the additional pressure for Gauss-Bonnet gravity, 
\begin{equation}\label{eq:pmg_GB}
p_{_{MG}} = \frac{n\left(n-1\right)\tilde{\alpha}}{\left(n-3\right)16\pi\ell_{p}^{n-1}\tilde{r}_{A}^2}\left[\frac{8\dot{\tilde{r}}_{A}}{H\tilde{r}_{A}^{3}}- \frac{\left(n+1\right)}{\tilde{r}_{A}^2}\right].
\end{equation}
Using the second Friedmann equation \ref{eq:Friedmanneqn2}, the additional effective pressure in Gauss-Bonnet gravity can be expressed as
\begin{equation}\label{eq:pmg_GB1}
p_{_{MG}} = \frac{4n\tilde{\alpha}}{\left(n-3\right)\tilde{r}_{A}^2 - \left(n+1\right)2\tilde{\alpha}}\left[\rho_{m} + p_{m}\right]- \rho_{_{MG}}.
\end{equation}
Then the $N_{bulk}$ in Gauss-Bonnet gravity will be
\begin{align}\label{eq:NbulkGB}
N_{bulk}&=-\epsilon \frac{E_{\text{Komar}}}{(1/2)T}\\
&=-\epsilon \frac{2\left[\left(n-2\right)\rho_{\text{eff}} + np_{\text{eff}}\right]V}{\left(n-1\right)(1/2)T},
\end{align}
where $\rho_{\text{eff}}$ and $ p_{\text{eff}} $ include additional components  $\rho_{\text{MG}}$ and $ p_{\text{MG}} $ from Equations (\ref{eq:rhomg_GB}) and (\ref{eq:pmg_GB1}), respectively, due to the modifications in gravity.

Now, from the unified expansion law in non-equilibrium, one can obtain the expansion law in Gauss-Bonnet gravity as
\begin{equation}\label{eq:expansionlaw_gaussBonnet}
\frac{dV}{dt}=\left(1 + \frac{n-1}{n-3}\frac{2\tilde{\alpha}}{\tilde{r}_{A}^{2}}\right)^{-1} \ell_{P}^{n-1}H\tilde{r}_{A}\left(N_{sur} -\epsilon N_{bulk}\right).
\end{equation}
where the $ N_{sur} $ and $ N_{bulk} $ are given by equations (\ref{eq:NsurGB}) and (\ref{eq:NbulkGB}), respectively. Putting the form of $G_{\text{eff}}$ from (\ref{eq:GBGeff}) and $ \rho_{\text{MG}} $  from (\ref{eq:rhomg_GB}), this expansion law can be easily reduced to Friedmann equation,
\begin{equation}\label{eq:FriedmannGauss-Bonnet}
H^2 +\frac{k}{a^2}+\tilde{\alpha}\left(H^2 +\frac{k}{a^2}\right)^{2} =\frac{16\pi \ell_{\!\!_p}^{n-1}}{n(n-1)}\rho_{m}.
\end{equation} 	

Newly obtained expansion law Gauss-Bonnet gravity (\ref{eq:expansionlaw_gaussBonnet})is different from the expansion law proposed in \cite{cai,Sheykhi2013}. The L.H.S. of our expansion law is exactly same as Padmanabhan's original proposal in \cite{paddy2012jun} and the surface degrees of freedom  follows the relation four times entropy,$ N_{sur}= 4S $. Another major difference is that the bulk degrees of freedom we have used in this paper depends on the effective Komar energy as
\begin{equation}\label{Effective MisnerSharpEnergyGB}
E = \left(1 + \frac{n-1}{n-3}\frac{2\tilde{\alpha}}{\tilde{r}_{A}^{2}}\right)\dfrac{n(n-1)\Omega_{n}\tilde{r}_{A}^{n-2}}{16\pi \ell_{P}^{n-1}},
\end{equation}
 which depends on the gravity theory. But in \cite{cai,Sheykhi2013} the bulk degrees of freedom is assumed as independent of gravity. Both the expansion laws are consistent with Friedmann equations in Gauss-Bonnet gravity, but the newly obtained expansion law seems more consistent with standard notions in emergent gravity paradigm.

 \subsection{Lovelock gravity}
 
For Lovelock gravity, the entropy has the form \cite{CAI2004237,Sheykhi2013},
\begin{equation}\label{eq:entropy_lovlock}
S = \dfrac{A}{4\ell_{P}^{n-1}}\sum_{i=1}^{m}\dfrac{i(n-1)}{(n-2i+1)}\hat{c}_{i}\tilde{r}_{_A}^{2-2i}.
\end{equation}
Then the surface degrees of freedom will takes the form,
\begin{equation}\label{eq:NsurLL}
N_{sur}  =\dfrac{A}{\ell_{P}^{n-1}}\sum_{i=1}^{m}\dfrac{i(n-1)}{(n-2i+1)}\hat{c}_{i}\tilde{r}_{_A}^{2-2i}.
\end{equation}
It can be noted that the density of surface degrees of freedom,$\sigma_{sur} $, for Lovelock gravity varies with the horizon radius, and the effective coupling strength can be obtained as	
\begin{equation}\label{eq:LLGeff}
G_{\text{eff}}=\left(\sigma_{sur}\right)^{-1} = \frac{\ell_{P}^{n-1}}{\sum_{i=1}^{m}\dfrac{i(n-1)}{(n-2i+1)}\hat{c}_{i}\tilde{r}_{_A}^{2-2i}}.
\end{equation}

Now the additional density component $ \rho_{_{MG}} $ due to the Lovelock gravity in the effective Misner-Sharp energy will be   
\begin{equation}\label{eq:rhomg_LL}
\rho_{_{MG}} = \frac{n(n-1)}{16\pi\ell_{P}^{n-1}}\sum_{i=1}^{m}   \frac{(n+1)(i-1)}{n+1-2i}\hat{c}_{i}\tilde{r}_{A}^{-2i}.
\end{equation}
Similarly, the additional pressure can be obtained as
\begin{align}\label{eq:pmg_LL}
p_{_{MG}} = \frac{n(n-1)}{16\pi\ell_{P}^{n-1}}\sum_{i=1}^{m}\frac{(1-i)\hat{c}_{i}}{(n+1-2i)}
\left[ \frac{(n+1)}{\tilde{r}_{A}^{2i}} 
 + \frac{4i\dot{\tilde{r}}_{_A}}{H\tilde{r}_{A}^{2i+1}} \right].
\end{align}
Using the second Friedmann equation \ref{eq:Friedmanneqn2}, the additional effective pressure in Lovelock gravity can be re-expressed as
\begin{align}\label{eq:pmg_LL1}
p_{_{MG}} =\frac{\Delta}{1-\Delta}\left(\rho_{m} +p_{m}\right) - \rho_{\text{MG}},
\end{align}
where the $ \Delta $ is 
\begin{align}
\Delta&=\\ &\sum_{i=1}^{m}\frac{2ni(1-i)\hat{c}_{i}}{(n+1-2i)\tilde{r}_{A}^{2i-2}}\frac{1}{\sum_{j=1}^{m}\frac{2nj(1-j)\hat{c}_{j}}{(n+1-2j)\tilde{r}_{A}^{2j-2}}}
\end{align}
Then the bulk degrees of freedom in Lovelock gravity will be
\begin{align}\label{eq:NbulkLL}
N_{bulk}&=-\epsilon \frac{E_{\text{Komar}}}{(1/2)T}\\
&=-\epsilon \frac{2\left[\left(n-2\right)\rho_{\text{eff}} + np_{\text{eff}}\right]V}{\left(n-1\right)(1/2)T},
\end{align}
where $\rho_{\text{eff}}$ and $ p_{\text{eff}} $ comprise additional components, represented by $\rho_{\text{MG}}$ in Eq. (\ref{eq:rhomg_LL}) and $ p_{\text{MG}} $ in Eq. (\ref{eq:pmg_LL1}), in addition to the normal matter components $\rho_{m}$ and $ p_{m} $ respectively.

Now one can obtain the expansion law in Lovelock gravity from the unified expansion law in non-equilibrium as
\begin{equation}
\frac{dV}{dt}=\frac{\ell_{P}^{n-1}H\tilde{r}_{A}}{\sum_{i=1}^{m}\dfrac{i(n-1)}{(n-2i+1)}\hat{c}_{i}\tilde{r}_{_A}^{2-2i}}\left(N_{sur} -\epsilon N_{bulk}\right),
\end{equation}
where the $ N_{sur} $ and $ N_{bulk} $ are taken from equations (\ref{eq:NsurLL}) and (\ref{eq:NbulkLL}) respectively. This expansion law can be easily reduced to equation (\ref{eq:Friedmanneqn1}), using the $G_{\text{eff}}$ (\ref{eq:LLGeff}) and $ \rho_{\text{MG}} $ (\ref{eq:rhomg_LL}) which leads to 
\begin{align} \label{eq: expansionlaw_lovlock}
\sum_{i=1}^{m} \left(\hat{c}_{i}\tilde{r}_{A}^{-2i}\right. & \left. - i\hat{c}_{i}\tilde{r}_{A}^{-2i-1}\dot{\tilde{r}}_{A}H^{-1}  \right)  = \nonumber\\ &- \frac{8\pi \ell_{P}^{n-1}}{n(n-1)}\left[(n-2)\rho + np \right].
\end{align}
From the above expansion law (\ref{eq: expansionlaw_lovlock}), the first Friedmann equations can be obtained, first by
multiplying both sides of the 
equations with the factor $2a\dot{a}$  
and then integrate the result using the continuity equation \cite{Sheykhi2013},
\begin{equation}\label{eq:FriedmannLovlock}
\sum_{i=1}^{m}\hat{c}_{i}\left(H^2 +\frac{k}{a^2}\right)^i =\frac{16\pi \ell_{\!\!_p}^{n-1}}{n(n-1)}\rho.
\end{equation} 

\subsection{f(R) gravity}
Now we derive the expansion law in $ f(R) $ gravity from the unified expansion law in non-equilibrium (\ref{eq:unified expansion law}). The $ f(R) $ gravity is one of the main candidates from the gravity models with higher-order curvature corrections which can explain the late acceleration of the Universe \cite{Nojiri-Odintsov:2011, Nojiri-Odintsov-vko:2017, Sotiriou:2008rp}.
The Einstein-Hilbert action of f(R) gravity has the form, 
\begin{equation}
A = \int d^{(n+1)}x\, \sqrt{-g}\left(f(R) + 2\kappa L_{m}\right),
\end{equation}
where $\kappa = 8\pi G $. From the variational principle, $\delta A = 0 $ we get the modified field equations which can also be re expressed as in Eq. (\ref{eq:field_equation}). The additional component in the effective energy momentum tensor in $f(R)$ gravity can be identified as 
\begin{align}
&T_{\mu\nu}^{(MG)} = \frac{1}{8\pi G}\times \nonumber\\&\left(\frac{f(R)-Rf^{\prime}(R)}{2}g_{\mu\nu}+\nabla_{\mu}\nabla_{\nu}f^{\prime}(R) -  g_{\mu\nu}\nabla^{2}f^{\prime}(R) \right).
\end{align}
Then the total effective energy-momentum tensor implies a effective density $\rho_{\text{eff}}$ and pressure $p_{\text{eff}}$ as
\begin{align} \label{eq:rho_eff}
\rho_{\text{eff}} = &\rho_{m} + \dfrac{1}{8\pi G}\left[\dfrac{ Rf^{\prime}(R)-f(R)}{2}- nH\dot{f}'(R)\right] \\ \label{eq: p_eff}
p_{\text{eff}} = &p_{m} + \dfrac{1}{8\pi G}\left[\dfrac{f(R) - Rf^{\prime}(R)}{2} + \ddot{f}'(R)\right. \nonumber\\ & ~~~~~~~~~~~~~~~~~~~~\quad \qquad + (n-1)H\dot{f}'(R)\bigg].
\end{align} 
The modified first Friedmann equation implies that the effective density $\rho_{\text{eff}}$ (\ref{eq:rho_eff}) will be couple with Einstein tensor, $G_{\mu\nu}$, and the coupling strength $G_{\text{eff}} = Gf^{\prime}(R)^{-1} $.

The Wald entropy for $f(R)$ gravity have the form \cite{PhysRevD.48.R3427}
\begin{equation}\label{eq:entropy_f(R)}
S= \dfrac{Af^{\prime}(R)}{4G}.
\end{equation}
The corresponding surface degrees of freedom will be
\begin{equation}\label{eq:Nsur_f(R)}
N_{sur}= \dfrac{Af^{\prime}(R)}{G}.
\end{equation}
Here on can note that the density of surface degrees of freedom in $ f(R) $ gravity depends on the $ f^{\prime}(R) $.  
The additional density and pressure in $f(R)$ gravity can be identified as 
\begin{equation}
\rho_{_{MG}} =  \dfrac{1}{8\pi G}\left[\dfrac{ Rf^{\prime}(R)-f(R)}{2}- nH\dot{f}'(R)\right],
\end{equation}
\begin{equation}
p_{_{MG}} =  \frac{1}{8\pi G}\left[\dfrac{f(R) - Rf^{\prime}(R)}{2} + \ddot{f}'(R)+ (n-1)H\dot{f}'(R)\right].
\end{equation}
Hence the bulk degrees of freedom in $ f(R) $ gravity can be expressed as
\begin{align}\label{eq:NbulkfR}
N_{bulk}=-\epsilon \frac{2\left[\left(n-2\right)\rho_{\text{eff}} + np_{\text{eff}}\right]V}{\left(n-1\right)(1/2)T}.
\end{align}

We have obtained the form of degrees of freedom on the surface and bulk, and $ \sigma_{sur} $ in $ f(R) $ gravity. Now, lets obtain the expansion law f(R) from the unified expansion law in non-equilibrium (\ref{eq:unified expansion law}),
\begin{equation}
\frac{dV}{dt}=f^{\prime}(R)^{-1} \ell_{P}^{n-1}H\tilde{r}_{A}\left(N_{sur} -\epsilon N_{bulk}\right)
\end{equation}

One can easily deduce the corresponding Friedmann equation in f(R) gravity from the expansion law as in Eq. (\ref{eq:Friedmanneqn1}).
Now, putting $G_{\text{eff}}$ and $ \rho_{\text{eff}} $, we get first Friedmann equation in $f(R)$ gravity,
\begin{align}\label{eq:1fieldeqn_f(R)}
H^{2}+\frac{k}{a^2}  =& \dfrac{16\pi G }{n(n-1)f^{\prime}(R)}\\ &~~\left[\rho_{m} +\dfrac{ Rf^{\prime}(R)-f(R)}{2}- nH\dot{f}'(R)\right]. 
\end{align}
From which, one can get the second Friedmann equation by taking differential of Eq. (\ref{eq:1fieldeqn_f(R)}), using the continuity equation (\ref{eq: Gcontiniutyeqn_non-equilibrium}), 
\begin{equation}\label{eq: 2fieldeqn_f(R)}
\dot{H} - \frac{k}{a^2} = -\dfrac{8\pi G}{(n-1)f^{\prime}(R)}\left[\rho_{m} + p_{m}  + \ddot{f}^{\prime}(R)-H\dot{f}^{\prime}(R)\right].
\end{equation}
Hence the expansion law is consistent with Friedmann equation in f(R) gravity \cite{AKBAR2007243}. 

Previously there have several attempt to obtain the expansion law in $ f(R) $ gravity, none of them was fully successful in obtaining an expansion law consistent with the Friedmann equations in $f(R)$ gravity \cite{PhysRevD.88.043518, Tu_2013, PhysRevD.88.084019}.
In reference \cite{Tu_2013}, the authors try to derive the expansion law 
in $f(R)$ gravity by taking $N_{sur}=4S = Af^{\prime}(R)/\ell_{P}^{n-1}$. 
However, the resulting dynamical equation seems erroneous and hence is not consistent with the standard Friedmann equation.
In extending this work, authors in \cite{PhysRevD.88.084019} 
generalized  $\dfrac{dV}{dt}$ in the expansion law with 
$\dfrac{\alpha}{(n-1)H}\dfrac{dN_{sur}}{dt}$ and proposed a dynamic equation (Eq.16 in the reference), which is valid only 
in 
equilibrium conditions, 
$\dot{f_{R}}=0$. Also, in the process of derivation, the authors seem to have omitted 
a term, $H\dot{f}'(R)/2f(R)$, with which 
it is impossible to arrive at the Friedmann equation in $f(R)$ gravity \cite{PhysRevD.88.084019}. In contrast to these, our approach generalizes the expansion in non-equilibrium conditions and  is consistent with the Friedmann equations.

%

\section{Conclusions and Discussions}\label{conclusion}
Padmanabhan has demonstrated that the expansion of our Universe can be understood as the emergence of space, where the dynamics are driven towards holographic equipartition, \( N_{sur} = N_{bulk} \). The rate of this emergence is governed by the difference in the degrees of freedom within the framework of Einstein gravity \cite{paddy2012jun}. This expansion law has attracted considerable attention, leading to various extensions and generalizations across different gravitational theories \cite{cai,Sheykhi2013,PhysRevD.88.043518, Tu_2013, PhysRevD.88.084019}. However, many of these extensions rely on ad hoc assumptions, prompting the need for a unified formulation applicable to a broader range of gravitational theories.

A promising attempt at such a unified formulation is presented in \cite{basari2}, which utilizes the first law of thermodynamics within an equilibrium framework. While this approach provides valuable insights into the expansion law and surface degrees of freedom in different gravity theories, it overlooks the non-equilibrium nature inherent in theories with higher-order curvature corrections. Consequently, the formulation encounters minor discrepancies. Specifically, it faces challenges in reconciling the fundamental relation of surface degrees of freedom on the surface, expressed as \( N_{sur} = 4S \), in general gravity theories such as Gauss-Bonnet and Lovelock gravity. Additionally, it struggles to derive the expansion law in gravity theories like \( f(R) \) gravity, which require a non-equilibrium treatment \cite{AKBAR2007243,PhysRevLett.96.121301,AKBAR20067}.

The discrepancy with the surface degrees of freedom in the unified expansion law is mainly due to the deviation of equipartition energy on the surface from \( 2TS \) in general theories of gravity. This challenge can be addressed by replacing the matter density with an effective density, ensuring the relation \( E_{eq} = \frac{4}{n-1} \rho_{eff} V = 2TS \). This can be achieved by choosing the total energy in the bulk as an effective Misner-Sharp energy as outlined in \cite{unifiedformulationTian:2014ila}. Then the resulting first law of thermodynamics naturally incorporates an additional entropy production term corresponding to non-equilibrium thermodynamics. Utilizing this modified first law with non-equilibrium considerations, we have developed a unified expansion law within a non-equilibrium framework. We have also shown that the Padmanabhan's original expansion law for (3+1) Einstein gravity in a flat FRW universe can be deduced from our unified expansion law in non-equilibrium. Furthermore, the Friedmann equations can also be derived from our unified expansion law in non-equilibrium for any minimally coupled gravity theories. These consistency underscores the robustness and reliability of our formulation.

One notable advantage of the unified expansion law is that the left-hand side of the equation, \(\frac{dV}{dt}\), precisely matches Padmanabhan's original proposal, even in the context of more general theories of gravity. This alignment helps avoid ad hoc assumptions when extending the original expansion law to more general theories of gravity. Furthermore, the unified expansion law in non-equilibrium reveals that the rate of emergence \(\frac{dV}{dt}\) is not only related to the difference in degrees of freedom \((N_{sur} - N_{bulk})\) but also inversely proportional to the density of surface degrees of freedom on the horizon. In more general theories of gravity, the density of surface degrees of freedom may vary over time. Hence, one can expect changes in the emergence of space; specifically, space will emerge more rapidly in situations where the density of surface degrees of freedom is lower.

The unified expansion law in non-equilibrium can be used to encompass a broader set of gravity theories. As examples, we have derived the expansion law in Gauss-Bonnet gravity, Lovelock gravity, and \(f(R)\) gravity, and compared our results with previous generalizations in the literature. The newly derived unified expansion law can also be applied to minimally coupled gravity theories such as Generalized Brans-Dicke gravity, scalar-tensor-chameleon gravity, quadratic gravity, and \(f(R, \mathcal{G})\) generalized Gauss-Bonnet gravity. This can be achieved by identifying the effective coupling strength, \(G_{\text{eff}}\), and the effective Misner-Sharp energy in these gravity theories, as demonstrated in \cite{unifiedformulationTian:2014ila}.

More importantly, in our formulation, the surface degrees of freedom consistently adhere to the essential relation \(N_{sur} = 4S\) across all minimally coupled gravity theories.  Additionally, the bulk degrees of freedom acquire an additional gravity component that depends on the specific gravity theory employed. Further exploration of the gravitational influences on bulk degrees of freedom may provide deeper insights into their true nature and significance.

\acknowledgments

I thank Titus K. Mathew and P.B. Krishna for fruitful discussions. I also acknowledge Cochin University of Science and Technology for the University Senior Research Fellowship.


\begin{thebibliography}{59}%
	\makeatletter
	\providecommand \@ifxundefined [1]{%
		\@ifx{#1\undefined}
	}%
	\providecommand \@ifnum [1]{%
		\ifnum #1\expandafter \@firstoftwo
		\else \expandafter \@secondoftwo
		\fi
	}%
	\providecommand \@ifx [1]{%
		\ifx #1\expandafter \@firstoftwo
		\else \expandafter \@secondoftwo
		\fi
	}%
	\providecommand \natexlab [1]{#1}%
	\providecommand \enquote  [1]{``#1''}%
	\providecommand \bibnamefont  [1]{#1}%
	\providecommand \bibfnamefont [1]{#1}%
	\providecommand \citenamefont [1]{#1}%
	\providecommand \href@noop [0]{\@secondoftwo}%
	\providecommand \href [0]{\begingroup \@sanitize@url \@href}%
	\providecommand \@href[1]{\@@startlink{#1}\@@href}%
	\providecommand \@@href[1]{\endgroup#1\@@endlink}%
	\providecommand \@sanitize@url [0]{\catcode `\\12\catcode `\$12\catcode
		`\&12\catcode `\#12\catcode `\^12\catcode `\_12\catcode `\%12\relax}%
	\providecommand \@@startlink[1]{}%
	\providecommand \@@endlink[0]{}%
	\providecommand \url  [0]{\begingroup\@sanitize@url \@url }%
	\providecommand \@url [1]{\endgroup\@href {#1}{\urlprefix }}%
	\providecommand \urlprefix  [0]{URL }%
	\providecommand \Eprint [0]{\href }%
	\providecommand \doibase [0]{http://dx.doi.org/}%
	\providecommand \selectlanguage [0]{\@gobble}%
	\providecommand \bibinfo  [0]{\@secondoftwo}%
	\providecommand \bibfield  [0]{\@secondoftwo}%
	\providecommand \translation [1]{[#1]}%
	\providecommand \BibitemOpen [0]{}%
	\providecommand \bibitemStop [0]{}%
	\providecommand \bibitemNoStop [0]{.\EOS\space}%
	\providecommand \EOS [0]{\spacefactor3000\relax}%
	\providecommand \BibitemShut  [1]{\csname bibitem#1\endcsname}%
	\let\auto@bib@innerbib\@empty
	\bibitem [{\citenamefont {Bekenstein}(1973)}]{bekenstein1}%
	\BibitemOpen
	\bibfield  {author} {\bibinfo {author} {\bibfnamefont {J.~D.}\ \bibnamefont
			{Bekenstein}},\ }\href {\doibase 10.1103/PhysRevD.7.2333} {\bibfield
		{journal} {\bibinfo  {journal} {Phys. Rev. D}\ }\textbf {\bibinfo {volume}
			{7}},\ \bibinfo {pages} {2333} (\bibinfo {year} {1973})}\BibitemShut
	{NoStop}%
	\bibitem [{\citenamefont {Bekenstein}(1974)}]{bekenstein2}%
	\BibitemOpen
	\bibfield  {author} {\bibinfo {author} {\bibfnamefont {J.~D.}\ \bibnamefont
			{Bekenstein}},\ }\href {\doibase 10.1103/PhysRevD.9.3292} {\bibfield
		{journal} {\bibinfo  {journal} {Phys. Rev. D}\ }\textbf {\bibinfo {volume}
			{9}},\ \bibinfo {pages} {3292} (\bibinfo {year} {1974})}\BibitemShut
	{NoStop}%
	\bibitem [{\citenamefont {Bardeen}\ \emph {et~al.}(1973)\citenamefont
		{Bardeen}, \citenamefont {Carter},\ and\ \citenamefont {Hawking}}]{bardeen}%
	\BibitemOpen
	\bibfield  {author} {\bibinfo {author} {\bibfnamefont {J.~M.}\ \bibnamefont
			{Bardeen}}, \bibinfo {author} {\bibfnamefont {B.}~\bibnamefont {Carter}}, \
		and\ \bibinfo {author} {\bibfnamefont {S.~W.}\ \bibnamefont {Hawking}},\
	}\href {\doibase 10.1007/BF01645742} {\bibfield  {journal} {\bibinfo
			{journal} {Communications in Mathematical Physics}\ }\textbf {\bibinfo
			{volume} {31}},\ \bibinfo {pages} {161} (\bibinfo {year} {1973})}\BibitemShut
	{NoStop}%
	\bibitem [{\citenamefont {Hawking}(1976)}]{Hawking1}%
	\BibitemOpen
	\bibfield  {author} {\bibinfo {author} {\bibfnamefont {S.~W.}\ \bibnamefont
			{Hawking}},\ }\href {\doibase 10.1103/PhysRevD.13.191} {\bibfield  {journal}
		{\bibinfo  {journal} {Phys. Rev. D}\ }\textbf {\bibinfo {volume} {13}},\
		\bibinfo {pages} {191} (\bibinfo {year} {1976})}\BibitemShut {NoStop}%
	\bibitem [{\citenamefont {Fulling}(1973)}]{PhysRevD.7.2850}%
	\BibitemOpen
	\bibfield  {author} {\bibinfo {author} {\bibfnamefont {S.~A.}\ \bibnamefont
			{Fulling}},\ }\href {\doibase 10.1103/PhysRevD.7.2850} {\bibfield  {journal}
		{\bibinfo  {journal} {Phys. Rev. D}\ }\textbf {\bibinfo {volume} {7}},\
		\bibinfo {pages} {2850} (\bibinfo {year} {1973})}\BibitemShut {NoStop}%
	\bibitem [{\citenamefont {Davies}(1975)}]{Davies_1975}%
	\BibitemOpen
	\bibfield  {author} {\bibinfo {author} {\bibfnamefont {P.~C.~W.}\
			\bibnamefont {Davies}},\ }\href {\doibase 10.1088/0305-4470/8/4/022}
	{\bibfield  {journal} {\bibinfo  {journal} {Journal of Physics A:
				Mathematical and General}\ }\textbf {\bibinfo {volume} {8}},\ \bibinfo
		{pages} {609} (\bibinfo {year} {1975})}\BibitemShut {NoStop}%
	\bibitem [{\citenamefont {Gibbons}\ and\ \citenamefont
		{Hawking}(1977)}]{PhysRevD.15.2738}%
	\BibitemOpen
	\bibfield  {author} {\bibinfo {author} {\bibfnamefont {G.~W.}\ \bibnamefont
			{Gibbons}}\ and\ \bibinfo {author} {\bibfnamefont {S.~W.}\ \bibnamefont
			{Hawking}},\ }\href {\doibase 10.1103/PhysRevD.15.2738} {\bibfield  {journal}
		{\bibinfo  {journal} {Phys. Rev. D}\ }\textbf {\bibinfo {volume} {15}},\
		\bibinfo {pages} {2738} (\bibinfo {year} {1977})}\BibitemShut {NoStop}%
	\bibitem [{\citenamefont {Unruh}(1976)}]{PhysRevD.14.870}%
	\BibitemOpen
	\bibfield  {author} {\bibinfo {author} {\bibfnamefont {W.~G.}\ \bibnamefont
			{Unruh}},\ }\href {\doibase 10.1103/PhysRevD.14.870} {\bibfield  {journal}
		{\bibinfo  {journal} {Phys. Rev. D}\ }\textbf {\bibinfo {volume} {14}},\
		\bibinfo {pages} {870} (\bibinfo {year} {1976})}\BibitemShut {NoStop}%
	\bibitem [{\citenamefont {Jacobson}(1995)}]{1995}%
	\BibitemOpen
	\bibfield  {author} {\bibinfo {author} {\bibfnamefont {T.}~\bibnamefont
			{Jacobson}},\ }\href {\doibase 10.1103/PhysRevLett.75.1260} {\bibfield
		{journal} {\bibinfo  {journal} {Phys. Rev. Lett.}\ }\textbf {\bibinfo
			{volume} {75}},\ \bibinfo {pages} {1260} (\bibinfo {year} {1995})},\ \Eprint
	{http://arxiv.org/abs/gr-qc/9504004} {arXiv:gr-qc/9504004 [gr-qc]}
	\BibitemShut {NoStop}%
	\bibitem [{\citenamefont {Cai}\ and\ \citenamefont {Kim}(2005)}]{Cai_Kim}%
	\BibitemOpen
	\bibfield  {author} {\bibinfo {author} {\bibfnamefont {R.-G.}\ \bibnamefont
			{Cai}}\ and\ \bibinfo {author} {\bibfnamefont {S.~P.}\ \bibnamefont {Kim}},\
	}\href {\doibase 10.1088/1126-6708/2005/02/050} {\bibfield  {journal}
		{\bibinfo  {journal} {JHEP}\ }\textbf {\bibinfo {volume} {02}},\ \bibinfo
		{pages} {050} (\bibinfo {year} {2005})},\ \Eprint
	{http://arxiv.org/abs/hep-th/0501055} {arXiv:hep-th/0501055} \BibitemShut
	{NoStop}%
	\bibitem [{\citenamefont {Padmanabhan}(2002)}]{Padmanabhan:2002sha}%
	\BibitemOpen
	\bibfield  {author} {\bibinfo {author} {\bibfnamefont {T.}~\bibnamefont
			{Padmanabhan}},\ }\href {\doibase 10.1088/0264-9381/19/21/306} {\bibfield
		{journal} {\bibinfo  {journal} {Class. Quant. Grav.}\ }\textbf {\bibinfo
			{volume} {19}},\ \bibinfo {pages} {5387} (\bibinfo {year} {2002})},\ \Eprint
	{http://arxiv.org/abs/gr-qc/0204019} {arXiv:gr-qc/0204019} \BibitemShut
	{NoStop}%
	\bibitem [{\citenamefont {Paranjape}\ \emph {et~al.}(2006)\citenamefont
		{Paranjape}, \citenamefont {Sarkar},\ and\ \citenamefont
		{Padmanabhan}}]{PhysRevD.74.104015}%
	\BibitemOpen
	\bibfield  {author} {\bibinfo {author} {\bibfnamefont {A.}~\bibnamefont
			{Paranjape}}, \bibinfo {author} {\bibfnamefont {S.}~\bibnamefont {Sarkar}}, \
		and\ \bibinfo {author} {\bibfnamefont {T.}~\bibnamefont {Padmanabhan}},\
	}\href {\doibase 10.1103/PhysRevD.74.104015} {\bibfield  {journal} {\bibinfo
			{journal} {Phys. Rev. D}\ }\textbf {\bibinfo {volume} {74}},\ \bibinfo
		{pages} {104015} (\bibinfo {year} {2006})}\BibitemShut {NoStop}%
	\bibitem [{\citenamefont {Akbar}\ and\ \citenamefont
		{Cai}(2007{\natexlab{a}})}]{PhysRevD.75.084003}%
	\BibitemOpen
	\bibfield  {author} {\bibinfo {author} {\bibfnamefont {M.}~\bibnamefont
			{Akbar}}\ and\ \bibinfo {author} {\bibfnamefont {R.-G.}\ \bibnamefont
			{Cai}},\ }\href {\doibase 10.1103/PhysRevD.75.084003} {\bibfield  {journal}
		{\bibinfo  {journal} {Phys. Rev. D}\ }\textbf {\bibinfo {volume} {75}},\
		\bibinfo {pages} {084003} (\bibinfo {year} {2007}{\natexlab{a}})}\BibitemShut
	{NoStop}%
	\bibitem [{\citenamefont {Hayward}(1998)}]{Hayward_1998}%
	\BibitemOpen
	\bibfield  {author} {\bibinfo {author} {\bibfnamefont {S.~A.}\ \bibnamefont
			{Hayward}},\ }\href {\doibase 10.1088/0264-9381/15/10/017} {\bibfield
		{journal} {\bibinfo  {journal} {Classical and Quantum Gravity}\ }\textbf
		{\bibinfo {volume} {15}},\ \bibinfo {pages} {3147} (\bibinfo {year}
		{1998})}\BibitemShut {NoStop}%
	\bibitem [{\citenamefont {Hayward}\ \emph {et~al.}(1999)\citenamefont
		{Hayward}, \citenamefont {Mukohyama},\ and\ \citenamefont
		{Ashworth}}]{Hayward:1998ee}%
	\BibitemOpen
	\bibfield  {author} {\bibinfo {author} {\bibfnamefont {S.~A.}\ \bibnamefont
			{Hayward}}, \bibinfo {author} {\bibfnamefont {S.}~\bibnamefont {Mukohyama}},
		\ and\ \bibinfo {author} {\bibfnamefont {M.~C.}\ \bibnamefont {Ashworth}},\
	}\href {\doibase 10.1016/S0375-9601(99)00225-X} {\bibfield  {journal}
		{\bibinfo  {journal} {Phys. Lett. A}\ }\textbf {\bibinfo {volume} {256}},\
		\bibinfo {pages} {347} (\bibinfo {year} {1999})},\ \Eprint
	{http://arxiv.org/abs/gr-qc/9810006} {arXiv:gr-qc/9810006} \BibitemShut
	{NoStop}%
	\bibitem [{\citenamefont {Kothawala}(2011)}]{PhysRevD.83.024026}%
	\BibitemOpen
	\bibfield  {author} {\bibinfo {author} {\bibfnamefont {D.}~\bibnamefont
			{Kothawala}},\ }\href {\doibase 10.1103/PhysRevD.83.024026} {\bibfield
		{journal} {\bibinfo  {journal} {Phys. Rev. D}\ }\textbf {\bibinfo {volume}
			{83}},\ \bibinfo {pages} {024026} (\bibinfo {year} {2011})}\BibitemShut
	{NoStop}%
	\bibitem [{\citenamefont
		{Padmanabhan}(2010{\natexlab{a}})}]{Padmanabhan:2009kr}%
	\BibitemOpen
	\bibfield  {author} {\bibinfo {author} {\bibfnamefont {T.}~\bibnamefont
			{Padmanabhan}},\ }\href {\doibase 10.1142/S021773231003313X} {\bibfield
		{journal} {\bibinfo  {journal} {Mod. Phys. Lett. A}\ }\textbf {\bibinfo
			{volume} {25}},\ \bibinfo {pages} {1129} (\bibinfo {year}
		{2010}{\natexlab{a}})},\ \Eprint {http://arxiv.org/abs/0912.3165}
	{arXiv:0912.3165 [gr-qc]} \BibitemShut {NoStop}%
	\bibitem [{\citenamefont {Padmanabhan}(2010{\natexlab{b}})}]{paddy2010dec}%
	\BibitemOpen
	\bibfield  {author} {\bibinfo {author} {\bibfnamefont {T.}~\bibnamefont
			{Padmanabhan}},\ }\href {\doibase 10.1142/S021773231003313X} {\bibfield
		{journal} {\bibinfo  {journal} {Mod. Phys. Lett.}\ }\textbf {\bibinfo
			{volume} {A25}},\ \bibinfo {pages} {1129} (\bibinfo {year}
		{2010}{\natexlab{b}})},\ \Eprint {http://arxiv.org/abs/0912.3165}
	{arXiv:0912.3165 [gr-qc]} \BibitemShut {NoStop}%
	\bibitem [{\citenamefont {Padmanabhan}(2019)}]{Padmanabhan2019review}%
	\BibitemOpen
	\bibfield  {author} {\bibinfo {author} {\bibfnamefont {T.}~\bibnamefont
			{Padmanabhan}},\ }\href {\doibase 10.1142/S0218271820300013} {\bibfield
		{journal} {\bibinfo  {journal} {Int. J. Mod. Phys. D}\ }\textbf {\bibinfo
			{volume} {29}},\ \bibinfo {pages} {2030001} (\bibinfo {year} {2019})},\
	\Eprint {http://arxiv.org/abs/1909.02015} {arXiv:1909.02015 [gr-qc]}
	\BibitemShut {NoStop}%
	\bibitem [{\citenamefont {{Padmanabhan}}(2012)}]{paddy2012jun}%
	\BibitemOpen
	\bibfield  {author} {\bibinfo {author} {\bibfnamefont {T.}~\bibnamefont
			{{Padmanabhan}}},\ }\href@noop {} {\bibfield  {journal} {\bibinfo  {journal}
			{arXiv e-prints}\ ,\ \bibinfo {eid} {arXiv:1206.4916}} (\bibinfo {year}
		{2012})},\ \Eprint {http://arxiv.org/abs/1206.4916} {arXiv:1206.4916
		[hep-th]} \BibitemShut {NoStop}%
	\bibitem [{\citenamefont {Cai}(2012)}]{cai}%
	\BibitemOpen
	\bibfield  {author} {\bibinfo {author} {\bibfnamefont {R.-G.}\ \bibnamefont
			{Cai}},\ }\href {\doibase 10.1007/JHEP11(2012)016} {\bibfield  {journal}
		{\bibinfo  {journal} {Journal of High Energy Physics}\ }\textbf {\bibinfo
			{volume} {2012}},\ \bibinfo {pages} {16} (\bibinfo {year}
		{2012})}\BibitemShut {NoStop}%
	\bibitem [{\citenamefont {Sheykhi}(2013)}]{Sheykhi2013}%
	\BibitemOpen
	\bibfield  {author} {\bibinfo {author} {\bibfnamefont {A.}~\bibnamefont
			{Sheykhi}},\ }\href {\doibase 10.1103/PhysRevD.87.061501} {\bibfield
		{journal} {\bibinfo  {journal} {Phys. Rev. D}\ }\textbf {\bibinfo {volume}
			{87}},\ \bibinfo {pages} {061501} (\bibinfo {year} {2013})}\BibitemShut
	{NoStop}%
	\bibitem [{\citenamefont {Padmanabhan}(2012)}]{paddycosmologicalconstant}%
	\BibitemOpen
	\bibfield  {author} {\bibinfo {author} {\bibfnamefont {T.}~\bibnamefont
			{Padmanabhan}},\ }\href@noop {} {\  (\bibinfo {year} {2012})},\ \Eprint
	{http://arxiv.org/abs/1210.4174} {arXiv:1210.4174 [hep-th]} \BibitemShut
	{NoStop}%
	\bibitem [{\citenamefont {Krishna}\ and\ \citenamefont
		{Mathew}(2017)}]{krishna1}%
	\BibitemOpen
	\bibfield  {author} {\bibinfo {author} {\bibfnamefont {P.~B.}\ \bibnamefont
			{Krishna}}\ and\ \bibinfo {author} {\bibfnamefont {T.~K.}\ \bibnamefont
			{Mathew}},\ }\href {\doibase 10.1103/PhysRevD.96.063513} {\bibfield
		{journal} {\bibinfo  {journal} {Phys. Rev. D}\ }\textbf {\bibinfo {volume}
			{96}},\ \bibinfo {pages} {063513} (\bibinfo {year} {2017})}\BibitemShut
	{NoStop}%
	\bibitem [{\citenamefont {Krishna}\ and\ \citenamefont
		{Mathew}(2019)}]{krishna2}%
	\BibitemOpen
	\bibfield  {author} {\bibinfo {author} {\bibfnamefont {P.~B.}\ \bibnamefont
			{Krishna}}\ and\ \bibinfo {author} {\bibfnamefont {T.~K.}\ \bibnamefont
			{Mathew}},\ }\href {\doibase 10.1103/PhysRevD.99.023535} {\bibfield
		{journal} {\bibinfo  {journal} {Phys. Rev.}\ }\textbf {\bibinfo {volume}
			{D99}},\ \bibinfo {pages} {023535} (\bibinfo {year} {2019})},\ \Eprint
	{http://arxiv.org/abs/1805.01705} {arXiv:1805.01705 [gr-qc]} \BibitemShut
	{NoStop}%
	\bibitem [{\citenamefont {Mahith}\ \emph {et~al.}(2018)\citenamefont {Mahith},
		\citenamefont {Krishna},\ and\ \citenamefont {Mathew}}]{Mahith2018}%
	\BibitemOpen
	\bibfield  {author} {\bibinfo {author} {\bibfnamefont {M.}~\bibnamefont
			{Mahith}}, \bibinfo {author} {\bibfnamefont {P.~B.}\ \bibnamefont {Krishna}},
		\ and\ \bibinfo {author} {\bibfnamefont {T.~K.}\ \bibnamefont {Mathew}},\
	}\href {\doibase 10.1088/1475-7516/2018/12/042} {\bibfield  {journal}
		{\bibinfo  {journal} {Journal of Cosmology and Astroparticle Physics}\
		}\textbf {\bibinfo {volume} {2018}},\ \bibinfo {pages} {042} (\bibinfo {year}
		{2018})}\BibitemShut {NoStop}%
	\bibitem [{\citenamefont {Hareesh}\ \emph {et~al.}(2019)\citenamefont
		{Hareesh}, \citenamefont {Krishna},\ and\ \citenamefont
		{Mathew}}]{Hareesh_2019}%
	\BibitemOpen
	\bibfield  {author} {\bibinfo {author} {\bibfnamefont {T.}~\bibnamefont
			{Hareesh}}, \bibinfo {author} {\bibfnamefont {P.~B.}\ \bibnamefont
			{Krishna}}, \ and\ \bibinfo {author} {\bibfnamefont {T.~K.}\ \bibnamefont
			{Mathew}},\ }\href {\doibase 10.1088/1475-7516/2019/12/024} {\bibfield
		{journal} {\bibinfo  {journal} {JCAP}\ }\textbf {\bibinfo {volume} {12}},\
		\bibinfo {pages} {024} (\bibinfo {year} {2019})},\ \Eprint
	{http://arxiv.org/abs/1908.03349} {arXiv:1908.03349 [gr-qc]} \BibitemShut
	{NoStop}%
	\bibitem [{\citenamefont {Muhsinath}\ \emph {et~al.}(2023)\citenamefont
		{Muhsinath}, \citenamefont {Basari V.~T.},\ and\ \citenamefont
		{Mathew}}]{Muhsinath}%
	\BibitemOpen
	\bibfield  {author} {\bibinfo {author} {\bibfnamefont {M.}~\bibnamefont
			{Muhsinath}}, \bibinfo {author} {\bibfnamefont {H.}~\bibnamefont {Basari
				V.~T.}}, \ and\ \bibinfo {author} {\bibfnamefont {T.~K.}\ \bibnamefont
			{Mathew}},\ }\href {\doibase 10.1007/s10714-023-03091-x} {\bibfield
		{journal} {\bibinfo  {journal} {Gen. Rel. Grav.}\ }\textbf {\bibinfo {volume}
			{55}},\ \bibinfo {pages} {43} (\bibinfo {year} {2023})},\ \Eprint
	{http://arxiv.org/abs/2211.01739} {arXiv:2211.01739 [gr-qc]} \BibitemShut
	{NoStop}%
	\bibitem [{\citenamefont {Hassan~Basari}\ \emph {et~al.}(2023)\citenamefont
		{Hassan~Basari}, \citenamefont {Krishna},\ and\ \citenamefont
		{Mathew}}]{basari2}%
	\BibitemOpen
	\bibfield  {author} {\bibinfo {author} {\bibfnamefont {V.~T.}\ \bibnamefont
			{Hassan~Basari}}, \bibinfo {author} {\bibfnamefont {P.~B.}\ \bibnamefont
			{Krishna}}, \ and\ \bibinfo {author} {\bibfnamefont {T.~K.}\ \bibnamefont
			{Mathew}},\ }\href {\doibase 10.1103/PhysRevD.107.063511} {\bibfield
		{journal} {\bibinfo  {journal} {Phys. Rev. D}\ }\textbf {\bibinfo {volume}
			{107}},\ \bibinfo {pages} {063511} (\bibinfo {year} {2023})}\BibitemShut
	{NoStop}%
	\bibitem [{\citenamefont {Komatsu}(2017)}]{Komatsu2016}%
	\BibitemOpen
	\bibfield  {author} {\bibinfo {author} {\bibfnamefont {N.}~\bibnamefont
			{Komatsu}},\ }\href {\doibase 10.1140/epjc/s10052-017-4800-2} {\bibfield
		{journal} {\bibinfo  {journal} {Eur. Phys. J.}\ }\textbf {\bibinfo {volume}
			{C77}},\ \bibinfo {pages} {229} (\bibinfo {year} {2017})},\ \Eprint
	{http://arxiv.org/abs/1611.04084} {arXiv:1611.04084 [gr-qc]} \BibitemShut
	{NoStop}%
	\bibitem [{\citenamefont {Ling}\ and\ \citenamefont
		{Pan}(2013)}]{PhysRevD.88.043518}%
	\BibitemOpen
	\bibfield  {author} {\bibinfo {author} {\bibfnamefont {Y.}~\bibnamefont
			{Ling}}\ and\ \bibinfo {author} {\bibfnamefont {W.-J.}\ \bibnamefont {Pan}},\
	}\href {\doibase 10.1103/PhysRevD.88.043518} {\bibfield  {journal} {\bibinfo
			{journal} {Phys. Rev. D}\ }\textbf {\bibinfo {volume} {88}},\ \bibinfo
		{pages} {043518} (\bibinfo {year} {2013})}\BibitemShut {NoStop}%
	\bibitem [{\citenamefont {Tu}\ and\ \citenamefont {Chen}(2013)}]{Tu_2013}%
	\BibitemOpen
	\bibfield  {author} {\bibinfo {author} {\bibfnamefont {F.-Q.}\ \bibnamefont
			{Tu}}\ and\ \bibinfo {author} {\bibfnamefont {Y.-X.}\ \bibnamefont {Chen}},\
	}\href {\doibase 10.1088/1475-7516/2013/05/024} {\bibfield  {journal}
		{\bibinfo  {journal} {Journal of Cosmology and Astroparticle Physics}\
		}\textbf {\bibinfo {volume} {2013}},\ \bibinfo {pages} {024} (\bibinfo {year}
		{2013})}\BibitemShut {NoStop}%
	\bibitem [{\citenamefont {Ai}\ \emph {et~al.}(2013)\citenamefont {Ai},
		\citenamefont {Hu}, \citenamefont {Chen},\ and\ \citenamefont
		{Deng}}]{PhysRevD.88.084019}%
	\BibitemOpen
	\bibfield  {author} {\bibinfo {author} {\bibfnamefont {W.-Y.}\ \bibnamefont
			{Ai}}, \bibinfo {author} {\bibfnamefont {X.-R.}\ \bibnamefont {Hu}}, \bibinfo
		{author} {\bibfnamefont {H.}~\bibnamefont {Chen}}, \ and\ \bibinfo {author}
		{\bibfnamefont {J.-B.}\ \bibnamefont {Deng}},\ }\href {\doibase
		10.1103/PhysRevD.88.084019} {\bibfield  {journal} {\bibinfo  {journal} {Phys.
				Rev. D}\ }\textbf {\bibinfo {volume} {88}},\ \bibinfo {pages} {084019}
		(\bibinfo {year} {2013})}\BibitemShut {NoStop}%
	\bibitem [{\citenamefont {Chakraborty}\ and\ \citenamefont
		{Padmanabhan}(2014)}]{PhysRevD.90.124017}%
	\BibitemOpen
	\bibfield  {author} {\bibinfo {author} {\bibfnamefont {S.}~\bibnamefont
			{Chakraborty}}\ and\ \bibinfo {author} {\bibfnamefont {T.}~\bibnamefont
			{Padmanabhan}},\ }\href {\doibase 10.1103/PhysRevD.90.124017} {\bibfield
		{journal} {\bibinfo  {journal} {Phys. Rev. D}\ }\textbf {\bibinfo {volume}
			{90}},\ \bibinfo {pages} {124017} (\bibinfo {year} {2014})}\BibitemShut
	{NoStop}%
	\bibitem [{\citenamefont {Sheykhi}(2018)}]{SHEYKHI2018118}%
	\BibitemOpen
	\bibfield  {author} {\bibinfo {author} {\bibfnamefont {A.}~\bibnamefont
			{Sheykhi}},\ }\href {\doibase https://doi.org/10.1016/j.physletb.2018.08.036}
	{\bibfield  {journal} {\bibinfo  {journal} {Physics Letters B}\ }\textbf
		{\bibinfo {volume} {785}},\ \bibinfo {pages} {118} (\bibinfo {year}
		{2018})}\BibitemShut {NoStop}%
	\bibitem [{\citenamefont {Tu}\ \emph {et~al.}(2018)\citenamefont {Tu},
		\citenamefont {Chen}, \citenamefont {Sun},\ and\ \citenamefont
		{Yang}}]{TU2018411}%
	\BibitemOpen
	\bibfield  {author} {\bibinfo {author} {\bibfnamefont {F.-Q.}\ \bibnamefont
			{Tu}}, \bibinfo {author} {\bibfnamefont {Y.-X.}\ \bibnamefont {Chen}},
		\bibinfo {author} {\bibfnamefont {B.}~\bibnamefont {Sun}}, \ and\ \bibinfo
		{author} {\bibfnamefont {Y.-C.}\ \bibnamefont {Yang}},\ }\href {\doibase
		https://doi.org/10.1016/j.physletb.2018.08.030} {\bibfield  {journal}
		{\bibinfo  {journal} {Physics Letters B}\ }\textbf {\bibinfo {volume}
			{784}},\ \bibinfo {pages} {411} (\bibinfo {year} {2018})}\BibitemShut
	{NoStop}%
	\bibitem [{\citenamefont {Komatsu}(2019)}]{PhysRevD.99.043523}%
	\BibitemOpen
	\bibfield  {author} {\bibinfo {author} {\bibfnamefont {N.}~\bibnamefont
			{Komatsu}},\ }\href {\doibase 10.1103/PhysRevD.99.043523} {\bibfield
		{journal} {\bibinfo  {journal} {Phys. Rev. D}\ }\textbf {\bibinfo {volume}
			{99}},\ \bibinfo {pages} {043523} (\bibinfo {year} {2019})}\BibitemShut
	{NoStop}%
	\bibitem [{\citenamefont {{Farag Ali}}(2014)}]{FARAGALI}%
	\BibitemOpen
	\bibfield  {author} {\bibinfo {author} {\bibfnamefont {A.}~\bibnamefont
			{{Farag Ali}}},\ }\href {\doibase
		https://doi.org/10.1016/j.physletb.2014.04.011} {\bibfield  {journal}
		{\bibinfo  {journal} {Physics Letters B}\ }\textbf {\bibinfo {volume}
			{732}},\ \bibinfo {pages} {335} (\bibinfo {year} {2014})}\BibitemShut
	{NoStop}%
	\bibitem [{\citenamefont {Yuan}\ and\ \citenamefont
		{Huang}(2017)}]{Yuan:2016pkz}%
	\BibitemOpen
	\bibfield  {author} {\bibinfo {author} {\bibfnamefont {F.-F.}\ \bibnamefont
			{Yuan}}\ and\ \bibinfo {author} {\bibfnamefont {P.}~\bibnamefont {Huang}},\
	}\href {\doibase 10.1088/1361-6382/aa61df} {\bibfield  {journal} {\bibinfo
			{journal} {Class. Quant. Grav.}\ }\textbf {\bibinfo {volume} {34}},\ \bibinfo
		{pages} {077001} (\bibinfo {year} {2017})},\ \Eprint
	{http://arxiv.org/abs/1607.04383} {arXiv:1607.04383 [gr-qc]} \BibitemShut
	{NoStop}%
	\bibitem [{\citenamefont {{V T}}\ \emph {et~al.}(2022)\citenamefont {{V T}},
		\citenamefont {{Krishna}}, \citenamefont {{K V}},\ and\ \citenamefont
		{{Mathew}}}]{basari1}%
	\BibitemOpen
	\bibfield  {author} {\bibinfo {author} {\bibfnamefont {H.~B.}\ \bibnamefont
			{{V T}}}, \bibinfo {author} {\bibfnamefont {P.~B.}\ \bibnamefont
			{{Krishna}}}, \bibinfo {author} {\bibfnamefont {P.}~\bibnamefont {{K V}}}, \
		and\ \bibinfo {author} {\bibfnamefont {T.~K.}\ \bibnamefont {{Mathew}}},\
	}\href {\doibase 10.1088/1361-6382/ac6a39} {\bibfield  {journal} {\bibinfo
			{journal} {Classical and Quantum Gravity}\ }\textbf {\bibinfo {volume}
			{39}},\ \bibinfo {eid} {115012} (\bibinfo {year} {2022})},\ \Eprint
	{http://arxiv.org/abs/1905.03552} {arXiv:1905.03552 [gr-qc]} \BibitemShut
	{NoStop}%
	\bibitem [{\citenamefont {Dheepika}\ \emph {et~al.}(2024)\citenamefont
		{Dheepika}, \citenamefont {T.},\ and\ \citenamefont
		{Mathew}}]{Dheepika:2022sio}%
	\BibitemOpen
	\bibfield  {author} {\bibinfo {author} {\bibfnamefont {M.}~\bibnamefont
			{Dheepika}}, \bibinfo {author} {\bibfnamefont {H.~B.~V.}\ \bibnamefont {T.}},
		\ and\ \bibinfo {author} {\bibfnamefont {T.~K.}\ \bibnamefont {Mathew}},\
	}\href {\doibase 10.1088/1402-4896/ad1375} {\bibfield  {journal} {\bibinfo
			{journal} {Phys. Scripta}\ }\textbf {\bibinfo {volume} {99}},\ \bibinfo
		{pages} {015014} (\bibinfo {year} {2024})},\ \Eprint
	{http://arxiv.org/abs/2211.14039} {arXiv:2211.14039 [gr-qc]} \BibitemShut
	{NoStop}%
	\bibitem [{\citenamefont {P}\ and\ \citenamefont {Mathew}(2023)}]{Nandu2023}%
	\BibitemOpen
	\bibfield  {author} {\bibinfo {author} {\bibfnamefont {N.~K.}\ \bibnamefont
			{P}}\ and\ \bibinfo {author} {\bibfnamefont {T.~K.}\ \bibnamefont {Mathew}},\
	}\href {\doibase 10.1016/j.dark.2023.101283} {\bibfield  {journal} {\bibinfo
			{journal} {Phys. Dark Univ.}\ }\textbf {\bibinfo {volume} {42}},\ \bibinfo
		{pages} {101283} (\bibinfo {year} {2023})},\ \Eprint
	{http://arxiv.org/abs/2302.01554} {arXiv:2302.01554 [gr-qc]} \BibitemShut
	{NoStop}%
	\bibitem [{\citenamefont {Krishna}\ \emph {et~al.}(2022)\citenamefont
		{Krishna}, \citenamefont {Hassan~Basari},\ and\ \citenamefont
		{Mathew}}]{Krishna2022}%
	\BibitemOpen
	\bibfield  {author} {\bibinfo {author} {\bibfnamefont {P.~B.}\ \bibnamefont
			{Krishna}}, \bibinfo {author} {\bibfnamefont {V.~T.}\ \bibnamefont
			{Hassan~Basari}}, \ and\ \bibinfo {author} {\bibfnamefont {T.~K.}\
			\bibnamefont {Mathew}},\ }\href {\doibase 10.1007/s10714-022-02941-4}
	{\bibfield  {journal} {\bibinfo  {journal} {General Relativity and
				Gravitation}\ }\textbf {\bibinfo {volume} {54}},\ \bibinfo {pages} {58}
		(\bibinfo {year} {2022})}\BibitemShut {NoStop}%
	\bibitem [{\citenamefont {Akbar}\ and\ \citenamefont
		{Cai}(2007{\natexlab{b}})}]{AKBAR2007243}%
	\BibitemOpen
	\bibfield  {author} {\bibinfo {author} {\bibfnamefont {M.}~\bibnamefont
			{Akbar}}\ and\ \bibinfo {author} {\bibfnamefont {R.-G.}\ \bibnamefont
			{Cai}},\ }\href {\doibase https://doi.org/10.1016/j.physletb.2007.03.005}
	{\bibfield  {journal} {\bibinfo  {journal} {Physics Letters B}\ }\textbf
		{\bibinfo {volume} {648}},\ \bibinfo {pages} {243} (\bibinfo {year}
		{2007}{\natexlab{b}})}\BibitemShut {NoStop}%
	\bibitem [{\citenamefont {Eling}\ \emph {et~al.}(2006)\citenamefont {Eling},
		\citenamefont {Guedens},\ and\ \citenamefont
		{Jacobson}}]{PhysRevLett.96.121301}%
	\BibitemOpen
	\bibfield  {author} {\bibinfo {author} {\bibfnamefont {C.}~\bibnamefont
			{Eling}}, \bibinfo {author} {\bibfnamefont {R.}~\bibnamefont {Guedens}}, \
		and\ \bibinfo {author} {\bibfnamefont {T.}~\bibnamefont {Jacobson}},\ }\href
	{\doibase 10.1103/PhysRevLett.96.121301} {\bibfield  {journal} {\bibinfo
			{journal} {Phys. Rev. Lett.}\ }\textbf {\bibinfo {volume} {96}},\ \bibinfo
		{pages} {121301} (\bibinfo {year} {2006})}\BibitemShut {NoStop}%
	\bibitem [{\citenamefont {Akbar}\ and\ \citenamefont
		{Cai}(2006{\natexlab{a}})}]{AKBAR20067}%
	\BibitemOpen
	\bibfield  {author} {\bibinfo {author} {\bibfnamefont {M.}~\bibnamefont
			{Akbar}}\ and\ \bibinfo {author} {\bibfnamefont {R.-G.}\ \bibnamefont
			{Cai}},\ }\href {\doibase https://doi.org/10.1016/j.physletb.2006.02.035}
	{\bibfield  {journal} {\bibinfo  {journal} {Physics Letters B}\ }\textbf
		{\bibinfo {volume} {635}},\ \bibinfo {pages} {7} (\bibinfo {year}
		{2006}{\natexlab{a}})}\BibitemShut {NoStop}%
	\bibitem [{\citenamefont {Hayward}(1996)}]{Hayward_1996}%
	\BibitemOpen
	\bibfield  {author} {\bibinfo {author} {\bibfnamefont {S.~A.}\ \bibnamefont
			{Hayward}},\ }\href {\doibase 10.1103/PhysRevD.53.1938} {\bibfield  {journal}
		{\bibinfo  {journal} {Phys. Rev. D}\ }\textbf {\bibinfo {volume} {53}},\
		\bibinfo {pages} {1938} (\bibinfo {year} {1996})}\BibitemShut {NoStop}%
	\bibitem [{\citenamefont {Tian}\ and\ \citenamefont
		{Booth}(2014{\natexlab{a}})}]{unifiedformulationTian:2014ila}%
	\BibitemOpen
	\bibfield  {author} {\bibinfo {author} {\bibfnamefont {D.~W.}\ \bibnamefont
			{Tian}}\ and\ \bibinfo {author} {\bibfnamefont {I.}~\bibnamefont {Booth}},\
	}\href {\doibase 10.1103/PhysRevD.90.104042} {\bibfield  {journal} {\bibinfo
			{journal} {Phys. Rev. D}\ }\textbf {\bibinfo {volume} {90}},\ \bibinfo
		{pages} {104042} (\bibinfo {year} {2014}{\natexlab{a}})},\ \Eprint
	{http://arxiv.org/abs/1409.4278} {arXiv:1409.4278 [gr-qc]} \BibitemShut
	{NoStop}%
	\bibitem [{\citenamefont {Tian}\ and\ \citenamefont
		{Booth}(2014{\natexlab{b}})}]{PhysRevD.90.104042}%
	\BibitemOpen
	\bibfield  {author} {\bibinfo {author} {\bibfnamefont {D.~W.}\ \bibnamefont
			{Tian}}\ and\ \bibinfo {author} {\bibfnamefont {I.}~\bibnamefont {Booth}},\
	}\href {\doibase 10.1103/PhysRevD.90.104042} {\bibfield  {journal} {\bibinfo
			{journal} {Phys. Rev. D}\ }\textbf {\bibinfo {volume} {90}},\ \bibinfo
		{pages} {104042} (\bibinfo {year} {2014}{\natexlab{b}})}\BibitemShut
	{NoStop}%
	\bibitem [{\citenamefont {Misner}\ and\ \citenamefont
		{Sharp}(1964)}]{PhysRev.136.B571}%
	\BibitemOpen
	\bibfield  {author} {\bibinfo {author} {\bibfnamefont {C.~W.}\ \bibnamefont
			{Misner}}\ and\ \bibinfo {author} {\bibfnamefont {D.~H.}\ \bibnamefont
			{Sharp}},\ }\href {\doibase 10.1103/PhysRev.136.B571} {\bibfield  {journal}
		{\bibinfo  {journal} {Phys. Rev.}\ }\textbf {\bibinfo {volume} {136}},\
		\bibinfo {pages} {B571} (\bibinfo {year} {1964})}\BibitemShut {NoStop}%
	\bibitem [{\citenamefont {Wald}(1993)}]{PhysRevD.48.R3427}%
	\BibitemOpen
	\bibfield  {author} {\bibinfo {author} {\bibfnamefont {R.~M.}\ \bibnamefont
			{Wald}},\ }\href {\doibase 10.1103/PhysRevD.48.R3427} {\bibfield  {journal}
		{\bibinfo  {journal} {Phys. Rev. D}\ }\textbf {\bibinfo {volume} {48}},\
		\bibinfo {pages} {R3427} (\bibinfo {year} {1993})}\BibitemShut {NoStop}%
	\bibitem [{\citenamefont {Jacobson}\ \emph {et~al.}(1994)\citenamefont
		{Jacobson}, \citenamefont {Kang},\ and\ \citenamefont
		{Myers}}]{Jacobson:1993}%
	\BibitemOpen
	\bibfield  {author} {\bibinfo {author} {\bibfnamefont {T.}~\bibnamefont
			{Jacobson}}, \bibinfo {author} {\bibfnamefont {G.}~\bibnamefont {Kang}}, \
		and\ \bibinfo {author} {\bibfnamefont {R.~C.}\ \bibnamefont {Myers}},\ }\href
	{\doibase 10.1103/PhysRevD.49.6587} {\bibfield  {journal} {\bibinfo
			{journal} {Phys. Rev. D}\ }\textbf {\bibinfo {volume} {49}},\ \bibinfo
		{pages} {6587} (\bibinfo {year} {1994})},\ \Eprint
	{http://arxiv.org/abs/gr-qc/9312023} {arXiv:gr-qc/9312023} \BibitemShut
	{NoStop}%
	\bibitem [{\citenamefont {Akbar}\ and\ \citenamefont
		{Cai}(2006{\natexlab{b}})}]{Akbar:2006er}%
	\BibitemOpen
	\bibfield  {author} {\bibinfo {author} {\bibfnamefont {M.}~\bibnamefont
			{Akbar}}\ and\ \bibinfo {author} {\bibfnamefont {R.-G.}\ \bibnamefont
			{Cai}},\ }\href {\doibase 10.1016/j.physletb.2006.02.035} {\bibfield
		{journal} {\bibinfo  {journal} {Phys. Lett. B}\ }\textbf {\bibinfo {volume}
			{635}},\ \bibinfo {pages} {7} (\bibinfo {year} {2006}{\natexlab{b}})},\
	\Eprint {http://arxiv.org/abs/hep-th/0602156} {arXiv:hep-th/0602156}
	\BibitemShut {NoStop}%
	\bibitem [{\citenamefont {Cai}(2002)}]{PhysRevD.65.084014}%
	\BibitemOpen
	\bibfield  {author} {\bibinfo {author} {\bibfnamefont {R.-G.}\ \bibnamefont
			{Cai}},\ }\href {\doibase 10.1103/PhysRevD.65.084014} {\bibfield  {journal}
		{\bibinfo  {journal} {Phys. Rev. D}\ }\textbf {\bibinfo {volume} {65}},\
		\bibinfo {pages} {084014} (\bibinfo {year} {2002})}\BibitemShut {NoStop}%
	\bibitem [{\citenamefont {Cai}\ and\ \citenamefont
		{Guo}(2004)}]{PhysRevD.69.104025}%
	\BibitemOpen
	\bibfield  {author} {\bibinfo {author} {\bibfnamefont {R.-G.}\ \bibnamefont
			{Cai}}\ and\ \bibinfo {author} {\bibfnamefont {Q.}~\bibnamefont {Guo}},\
	}\href {\doibase 10.1103/PhysRevD.69.104025} {\bibfield  {journal} {\bibinfo
			{journal} {Phys. Rev. D}\ }\textbf {\bibinfo {volume} {69}},\ \bibinfo
		{pages} {104025} (\bibinfo {year} {2004})}\BibitemShut {NoStop}%
	\bibitem [{\citenamefont {Cai}(2004)}]{CAI2004237}%
	\BibitemOpen
	\bibfield  {author} {\bibinfo {author} {\bibfnamefont {R.-G.}\ \bibnamefont
			{Cai}},\ }\href {\doibase https://doi.org/10.1016/j.physletb.2004.01.015}
	{\bibfield  {journal} {\bibinfo  {journal} {Physics Letters B}\ }\textbf
		{\bibinfo {volume} {582}},\ \bibinfo {pages} {237} (\bibinfo {year}
		{2004})}\BibitemShut {NoStop}%
	\bibitem [{\citenamefont {Nojiri}\ and\ \citenamefont
		{Odintsov}(2011)}]{Nojiri-Odintsov:2011}%
	\BibitemOpen
	\bibfield  {author} {\bibinfo {author} {\bibfnamefont {S.}~\bibnamefont
			{Nojiri}}\ and\ \bibinfo {author} {\bibfnamefont {S.~D.}\ \bibnamefont
			{Odintsov}},\ }\href {\doibase 10.1016/j.physrep.2011.04.001} {\bibfield
		{journal} {\bibinfo  {journal} {Phys. Rept.}\ }\textbf {\bibinfo {volume}
			{505}},\ \bibinfo {pages} {59} (\bibinfo {year} {2011})},\ \Eprint
	{http://arxiv.org/abs/1011.0544} {arXiv:1011.0544 [gr-qc]} \BibitemShut
	{NoStop}%
	\bibitem [{\citenamefont {Nojiri}\ \emph {et~al.}(2017)\citenamefont {Nojiri},
		\citenamefont {Odintsov},\ and\ \citenamefont
		{Oikonomou}}]{Nojiri-Odintsov-vko:2017}%
	\BibitemOpen
	\bibfield  {author} {\bibinfo {author} {\bibfnamefont {S.}~\bibnamefont
			{Nojiri}}, \bibinfo {author} {\bibfnamefont {S.~D.}\ \bibnamefont
			{Odintsov}}, \ and\ \bibinfo {author} {\bibfnamefont {V.~K.}\ \bibnamefont
			{Oikonomou}},\ }\href {\doibase 10.1016/j.physrep.2017.06.001} {\bibfield
		{journal} {\bibinfo  {journal} {Phys. Rept.}\ }\textbf {\bibinfo {volume}
			{692}},\ \bibinfo {pages} {1} (\bibinfo {year} {2017})},\ \Eprint
	{http://arxiv.org/abs/1705.11098} {arXiv:1705.11098 [gr-qc]} \BibitemShut
	{NoStop}%
	\bibitem [{\citenamefont {Sotiriou}\ and\ \citenamefont
		{Faraoni}(2010)}]{Sotiriou:2008rp}%
	\BibitemOpen
	\bibfield  {author} {\bibinfo {author} {\bibfnamefont {T.~P.}\ \bibnamefont
			{Sotiriou}}\ and\ \bibinfo {author} {\bibfnamefont {V.}~\bibnamefont
			{Faraoni}},\ }\href {\doibase 10.1103/RevModPhys.82.451} {\bibfield
		{journal} {\bibinfo  {journal} {Rev. Mod. Phys.}\ }\textbf {\bibinfo {volume}
			{82}},\ \bibinfo {pages} {451} (\bibinfo {year} {2010})},\ \Eprint
	{http://arxiv.org/abs/0805.1726} {arXiv:0805.1726 [gr-qc]} \BibitemShut
	{NoStop}%
\end{thebibliography}
%

\end{document}